\documentclass[twocolumn]{aastex631}

\usepackage{algorithm}
\usepackage{algpseudocode}
\usepackage{silence}
\usepackage{float}
\usepackage{tabularx}
\usepackage{multirow}
\usepackage{amsmath}

\shorttitle{Irregularly Sampled Time Series Interpolation for Binary Evolution Simulations}
\shortauthors{M. Srivastava et al.}
\graphicspath{{./}{figures/}}

\begin{document}

\title{Irregularly Sampled Time Series Interpolation for Detailed Binary Evolution Simulations}

\author[0000-0003-1749-6295]{Philipp\,M.\,Srivastava}
\affiliation{Electrical and Computer Engineering, Northwestern University, 2145 Sheridan Road, Evanston, IL 60208, USA}
\affiliation{Center for Interdisciplinary Exploration and Research in Astrophysics (CIERA), Northwestern University, 1800 Sherman Ave, Evanston, IL 60201, USA}
\affiliation{NSF-Simons AI Institute for the Sky (SkAI),172 E. Chestnut St., Chicago, IL 60611, USA}

\author[0000-0001-7273-0211]{Ugur\,Demir}
\affiliation{Electrical and Computer Engineering, Northwestern University, 2145 Sheridan Road, Evanston, IL 60208, USA}
\affiliation{Center for Interdisciplinary Exploration and Research in Astrophysics (CIERA), Northwestern University, 1800 Sherman Ave, Evanston, IL 60201, USA}
\affiliation{NSF-Simons AI Institute for the Sky (SkAI),172 E. Chestnut St., Chicago, IL 60611, USA}

\author[0000-0003-4554-0070]{Aggelos\,Katsaggelos}
\affiliation{Electrical and Computer Engineering, Northwestern University, 2145 Sheridan Road, Evanston, IL 60208, USA}
\affiliation{Center for Interdisciplinary Exploration and Research in Astrophysics (CIERA), Northwestern University, 1800 Sherman Ave, Evanston, IL 60201, USA}
\affiliation{NSF-Simons AI Institute for the Sky (SkAI),172 E. Chestnut St., Chicago, IL 60611, USA}

\author[0000-0001-9236-5469]{Vicky\,Kalogera}
\affiliation{Department of Physics and Astronomy, Northwestern University, 2145 Sheridan Road, Evanston, IL 60208, USA}
\affiliation{Center for Interdisciplinary Exploration and Research in Astrophysics (CIERA), Northwestern University, 1800 Sherman Ave, Evanston, IL 60201, USA}
\affiliation{NSF-Simons AI Institute for the Sky (SkAI),172 E. Chestnut St., Chicago, IL 60611, USA}

\author[0000-0002-2677-8019]{Shamal\,Lalvani}
\affiliation{Electrical and Computer Engineering, Northwestern University, 2145 Sheridan Road, Evanston, IL 60208, USA}

\author[0000-0003-0420-2067]{Elizabeth\,Teng}
\affiliation{Department of Physics and Astronomy, Northwestern University, 2145 Sheridan Road, Evanston, IL 60208, USA}
\affiliation{Center for Interdisciplinary Exploration and Research in Astrophysics (CIERA), Northwestern University, 1800 Sherman Ave, Evanston, IL 60201, USA}
\affiliation{NSF-Simons AI Institute for the Sky (SkAI),172 E. Chestnut St., Chicago, IL 60611, USA}

\author[0000-0003-1474-1523]{Tassos\,Fragos}
\affiliation{Département d’Astronomie, Université de Genève, Chemin Pegasi 51, CH-1290 Versoix, Switzerland}
\affiliation{Gravitational Wave Science Center (GWSC), Université de Genève, CH1211 Geneva, Switzerland}

\author[0000-0001-5261-3923]{Jeff\, J.\,Andrews}
\affiliation{Department of Physics, University of Florida, 2001 Museum Rd, Gainesville, FL 32611, USA}
\affiliation{Institute for Fundamental Theory, 2001 Museum Rd, Gainesville, FL 32611, USA}

\author[0000-0002-3439-0321]{Simone\,S.\,Bavera}
\affiliation{Département d’Astronomie, Université de Genève, Chemin Pegasi 51, CH-1290 Versoix, Switzerland}
\affiliation{Gravitational Wave Science Center (GWSC), Université de Genève, CH1211 Geneva, Switzerland}

\author[0000-0002-6842-3021]{Max\,Briel}
\affiliation{Département d’Astronomie, Université de Genève, Chemin Pegasi 51, CH-1290 Versoix, Switzerland}
\affiliation{Gravitational Wave Science Center (GWSC), Université de Genève, CH1211 Geneva, Switzerland}

\author[0000-0001-6692-6410]{Seth\,Gossage}
\affiliation{Center for Interdisciplinary Exploration and Research in Astrophysics (CIERA), Northwestern University, 1800 Sherman Ave, Evanston, IL 60201, USA}
\affiliation{NSF-Simons AI Institute for the Sky (SkAI),172 E. Chestnut St., Chicago, IL 60611, USA}

\author[0000-0003-3684-964X]{Konstantinos\,Kovlakas}
\affiliation{Institute of Space Sciences (ICE, CSIC), Campus UAB, Carrer de Magrans, 08193 Barcelona, Spain}
\affiliation{Institut d'Estudis Espacials de Catalunya (IEEC),  Edifici RDIT, Campus UPC, 08860 Castelldefels (Barcelona), Spain}

\author[0000-0001-9331-0400]{Matthias\,U.\,Kruckow}
\affiliation{Département d’Astronomie, Université de Genève, Chemin Pegasi 51, CH-1290 Versoix, Switzerland}
\affiliation{Gravitational Wave Science Center (GWSC), Université de Genève, CH1211 Geneva, Switzerland}

\author[0000-0002-8883-3351]{Camille\,Liotine}
\affiliation{Department of Physics and Astronomy, Northwestern University, 2145 Sheridan Road, Evanston, IL 60208, USA}
\affiliation{Center for Interdisciplinary Exploration and Research in Astrophysics (CIERA), Northwestern University, 1800 Sherman Ave, Evanston, IL 60201, USA}

\author[0000-0003-4474-6528]{Kyle\,A.\,Rocha}
\affiliation{Department of Physics and Astronomy, Northwestern University, 2145 Sheridan Road, Evanston, IL 60208, USA}
\affiliation{Center for Interdisciplinary Exploration and Research in Astrophysics (CIERA), Northwestern University, 1800 Sherman Ave, Evanston, IL 60201, USA}
\affiliation{NSF-Simons AI Institute for the Sky (SkAI),172 E. Chestnut St., Chicago, IL 60611, USA}

\author[0000-0001-9037-6180]{Meng\,Sun}
\affiliation{Center for Interdisciplinary Exploration and Research in Astrophysics (CIERA), Northwestern University, 1800 Sherman Ave, Evanston, IL 60201, USA}

\author[0000-0002-0031-3029]{Zepei\,Xing}
\affiliation{Département d’Astronomie, Université de Genève, Chemin Pegasi 51, CH-1290 Versoix, Switzerland}
\affiliation{Gravitational Wave Science Center (GWSC), Université de Genève, CH1211 Geneva, Switzerland}

\author[0000-0002-7464-498X]{Emmanouil\,Zapartas}
\affiliation{Institute of Astrophysics, Foundation for Research and Technology-Hellas, GR-71110 Heraklion, Greece}



\begin{abstract}

Modeling of large populations of binary stellar systems is an intergral part of a many areas of astrophysics, from radio pulsars and supernovae to X-ray binaries, gamma-ray bursts, and gravitational-wave mergers. Binary population synthesis codes that employ self-consistently the most advanced physics treatment available for stellar interiors and their evolution and are at the same time computationally tractable have started to emerge only recently. One element that is still missing from these codes is the ability to generate the complete time evolution of binaries with arbitrary initial conditions using pre-computed three-dimensional grids of binary sequences. Here we present a highly interpretable method, from binary evolution track interpolation. Our method implements simulation generation from irregularly sampled time series. Our results indicate that this method is appropriate for applications within binary population synthesis and computational astrophysics with time-dependent simulations in general. Furthermore we point out and offer solutions to the difficulty surrounding evaluating performance of signals exhibiting extreme morphologies akin to discontinuities.

\end{abstract}



\section{Introduction} \label{sec:intro}

Understanding the properties of binary-star systems is crucial to many disciplines within astronomy. Thus simulating their evolution is a major area of interest, especially through interactive phases of mass and angular momentum transfer and losses from the system. Most commonly for individual or small numbers of systems binary evolution is simulated using full stellar structure codes that account for the evolution of both binary components and their orbital properties, such as the open-source code {\tt MESA}  \citep{2011ApJS..192....3P, 2013ApJS..208....4P,2015ApJS..220...15P, 2018ApJS..234...34P,2019ApJS..243...10P,2023ApJS..265...15J}. These simulations involve solving differential equations on adaptive stellar interior grids, and although they are one-dimensional, they can take $1-100$ CPU hours  \citep{paxton_modules_2019} for one binary evolution sequence. Often studies of whole populations of binaries ($\gtrsim\,10^6$ binaries) are required (e.g., for comparisons to observed samples, models for whole galaxies or cosmological models, and relevant predictions) rendering the straightforward use of stellar structure and evolution codes in population studies very challenging if not impossible. 

In the past $2-3$ decades the problem of binary population synthesis has been addressed with the development and use of rapid synthesis codes which employ fitting formulas capturing the evolution of single stars \citep[][calculated with full stellar structure and evolution codes]{1995MNRAS.274..964P, 1997MNRAS.291..732T, 2000MNRAS.315..543H}, wrapping them with prescriptions that model the effects of binary interactions \citep[e.g.,][]{2001A&A...365..491N, 2002MNRAS.329..897H, 2008ApJS..174..223B, 2012A&A...546A..70T, 2018MNRAS.480.2011G, 2018MNRAS.481.1908K, 2019MNRAS.485..889S, 2020ApJ...898...71B, 2021arXiv210910352T}. 
Two population synthesis codes have incorporated full stellar structure and evolution codes for the modeling of interacting binary sequences: {\tt BPASS} \citep{2017PASA...34...58E, 2018MNRAS.479...75S}, for one of the binary components, and most recently {\tt POSYDON} \citep[][Andrews et al.\ 2024, v1 and v2, respectively]{fragos_posydon_2023} for both binary components. 
Although the development of these two population codes differs in many ways, in both cases large-scale grids of binary evolution sequences (tracks) are used. The generation of model populations then requires the 
prediction of tracks at arbitrary sub-grid points.  One way to carry out this prediction is through some kind of simple or more advanced ``interpolation'' through these grids, even if the existing tracks in the grid are used to assign properties to the binaries in the population. 

In {\tt POSYDON} v1 and v2 we incorporate three 3D different grids of interacting binary sequences modeled with {\tt MESA} to cover different evolutionary phases throughout the lives of binaries. They involve binaries that start with two H-rich Main Sequence stars, or with one compact object (neutron star or black hole) and either a H-rich or a He-rich Main Sequence star. In both v1 and v2 we incorporate advanced (improved in v2) methods for interpolating the properties of the binaries at the end of each grid phase based on their initial grid properties and the information from the pre-calculated grid tracks. This type of initial-final interpolation enables a wide range of population studies (many already appeared in the literature) but not those that require knowledge of the binaries' properties as a function of time at any arbitrary time point. For the latter, what is required is full binary track interpolation inside each of our 3D grids. 

In this paper we address the challenge of producing reliable full track interpolation using 3D binary evolution grids for any binaries with arbitrary properties that do not correspond to those in the pre-calculated grids. We note that this track-interpolation problem has been solved in the case of single stars (1D interpolation) using different methods akin to age proxies  \citep[][described in Section 3.2]{dotter_mesa_2016, maltsev_scalable_2024}. 
However, as binary evolution tracks exhibit highly complex morphologies (including Dirac delta-like changes), the single-star methods are rendered ineffective. In what follows we describe the formal problem of binary-evolution track interpolation in Section 2 and discuss its challenges in the context of existing work in Section 3. We present our new method in Section 4, analyze results from several evaluation tests in Section 5, and close with our Conclusions in Section 6.

\section{Problem Definition}

In this section we formally define the binary-evolution track interpolation problem with respect to a single grid which corresponds to any of the three such grids in the {\tt POSYDON} infrastructure:   the Hydrogen Main Sequence - Hydrogen Main Sequence (HMS-HMS), Compact Object - Hydrogen Main Sequence in Roche Lobe Overflow (CO-HMS), or, Compact Object - Helium Main Sequence (CO-HeMS) grids.  
We define $G$ as

\begin{equation}
    G = \{ \mathbf{i}_1, \mathbf{i}_2, \mathbf{i}_3, .., \mathbf{i}_N\},
\end{equation}

\noindent where $\mathbf{i}_j = (M_1^j, M_2^j, P^j)$ and $M_1^j$, $M_2^j$, and $P^j$ represent the initial mass of the primary star, the initial mass of the secondary star, and the initial orbital period of the star system, respectively, at the $j$th grid point. Each triplet represents the initial conditions of a given binary evolution track (``signal'') in the grid. For the CO-HMS RLO and CO-HeMS grids each of these quantities is uniformly varied in $\log_{10}$ space, while in the HMS-HMS grid, $M_2^j / M_1^j$ is varied (instead of $M_2^j$ in $\log_{10}$ space). Thus, the CO-HMS RLO and CO-HeMS grids form a three dimensional regular grid, and the HMS-HMS grid includes only gridpoints where $M_1^j > M_2^j$. The grid points are ordered in a certain way (e.g., lexicographically) to form the set $G$. Each element of $G$ defines the initial conditions of a simulated track $S_{p}(\mathbf{i}_j)$, that is, 
\begin{equation}
    S_{p}(\mathbf{i}_j) = \{ v_p(a_1^j), v_p(a_2^j), ... v_p(a_{F(\mathbf{i}_j)}^j) \}
    \label{equation:simulation}
\end{equation}

\noindent
where $a_n^j$ denotes the age of the star (measured in years) at timestep $n$, $v_p(a_i^j)$ denotes the value of the parameter $p$ at time $a_n^j$, and $F(\mathbf{i}_j)$ denotes the total number of timesteps in the simulated track at the grid point $\mathbf{i}_j$. Note that the temporal (age) distance between timesteps is not constant, that is the signals are not uniformly sampled.

\begin{figure*}
    \centering
    \includegraphics[width=\textwidth]{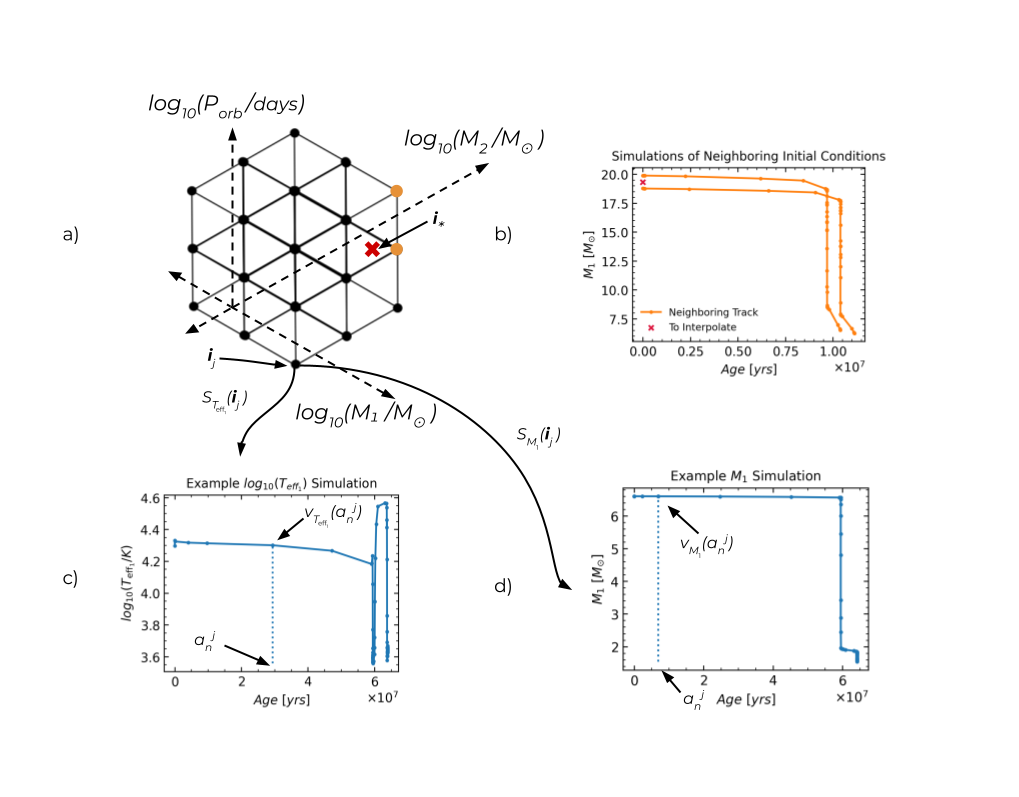}
    \caption{Panel a) shows an example grid $G$. The red marker represents the initial conditions to a simulation we wish to estimate. Panel b) shows the simulations of $M_1$ over time at the nearest initial conditions that we want to predict. It should be noted that we may use any number of simulations we may use any number of simulations to construct an approximate simulation at the initial conditions indicated by the red marker. Panels c) and d) show simulations corresponding to grid point $\mathbf{i}_j$. Panel c) shows an example simulation of the $log_{10}(T_\mathrm{eff, 1})$ parameter plotted against time, while panel d) shows an example simulation of the $M_1$ parameter also plotted against time.}
    \label{fig:problem_description}
\end{figure*}

The problem then becomes predicting $S_{p}(\mathbf{i}_*)$ for some $\mathbf{i}_* = (M_1^*, M_2^*, P^*)$ not included in $G$. A visual description of the problem can be seen in Figure \ref{fig:problem_description} where the grid in panel a) shows $G$ in $log_{10}$ space, and panel b) shows two simulations in orange which have similar initial conditions to some example point of interest denoted by the red marker. Panels c) and d) show example simulations for $\log_{10}(T_{\mathrm{eff, 1}})$ and $M_1$ respectively.

In this paper we will consider the generation of interpolated tracks for 10 different parameters $p$:

\begin{enumerate}
    \item $M_1$: the mass of the primary star measured in solar masses which is typically characterized as a signal with very little change without strong stellar winds or binary interactions. 
    \item $M_2$: the mass of the secondary star also measured in solar masses which is similar to $M_1$ in that very little change occurs without strong stellar winds or binary interactions.
    \item $P$: the orbital period of the system which is measured in days. Aggressive changes in signal value are also attributed to a wide range of binary interactions.
    \item $\log_{10}(\dot{M}_{\mathrm{transfer}})$: the $\log_{10}$ mass transfer rate in solar masses per year, which is typically either constant if no mass transfer occurs or is characterized by a Dirac delta-like signal when mass transfer occurs because mass-transfer phases typically last for very short time intervals compared to the full duration of the signal. Note that because of this special nature of the signal a special treatment will be given to it which is described in the proposed method section.
    \item $\log_{10}(T_{\mathrm{eff, 1}})$ and $\log_{10}(T_{\mathrm{eff, 2}})$: the $\log_{10}$ effective temperature of the primary and secondary star measured in Kelvin.
    \item $\log_{10}(L_1)$ and $\log_{10}(L_2)$: the $\log_{10}$ luminosity of the primary and secondary star measured in solar luminosity.
    \item  $\log_{10}(R_1)$ and $\log_{10}(R_2)$: the $\log_{10}$ radius of the primary or secondary star measured in solar radii.
\end{enumerate}

\noindent Note that quantities pertaining to the secondary star which are unique to stars ($L$, $T_{\mathrm{eff}}$, and $R$) are only applicable to the HMS-HMS grid.

\section{Challenges and Related Work}

\subsection{Challenges of the Problem}

In signal processing the standard procedure for non-uniformly sampled signals is to resample them at the Nyquist-Shannon rate \citep{oppenheim1999discrete}. However, the simulations at hand typically exhibit large changes in signal value over relatively short periods of time, that is, the signal includes high-frequency content. Thus, resampling them at the Nyquist-Shannon rate would result in a very small sampling period, making their processing and storage prohibitively demanding. We therefore do not consider interpolation methods in the existing literature which assume uniform sampling. 

Furthermore, the binary-star simulation tracks exhibit the \textit{timescale variance problem} \citep{maltsev_scalable_2024}. This refers to the fact that simulations exhibit certain morphologies for different amounts of time depending on the initial conditions. For single stars, the astrophysics of this behavior is well understood and is tractable, given initial conditions, and therefore one can anticipate the points in the tracks where morphology changes are to be expected. However, for interacting binaries, not only are the morphologies of the tracks themselves a lot more complex, but the triggers for morphological changes are not easy to identify astrophysically a priori based on the initial conditions. Therefore, it is imperative to be able to find the points in the tracks where significant morphology changes occur. 
The method we present here addresses this problem by identifying such points of significant change in signal morphology and by aligning them across tracks in a physically meaningful way, enabling effective interpolation. 

\subsection{Track Interpolation Problems in Astrophysics}

In the astrophysics literature solving the track interpolation problem for simple stellar models has been carried out in primarily two ways: using Equivalent Evolutionary Phases (EEPs) \citep{dotter_mesa_2016} and an age proxy \citep{maltsev_scalable_2024}. Both of these approaches involve simulations where the initial mass of the system is varied. In the first approach, astrophysically significant points known as primary EEPs are placed at changes in stellar evolution phase. Then, the secondary EEP locations are found by equally spacing a predetermined number of points along the simulation. Specifically the equal spacing refers to spacing on the Hertzsprung-Russel-diagram (HR-diagram) \citep{hertzsprung_zur_1905, russel_problems_1919} of the star. The effect of this is that all simulations are represented by the same number of time steps. Thus, track interpolation can be carried out by taking the weighted average of tracks with adjacent initial stellar masses (to the initial stellar mass of interest) \cite{dotter_mesa_2016}. While the second approach \citep{maltsev_scalable_2024} is considerably different, it uses similar ideas to solve the timescale variance problem. Instead of interpolating one track at once they interpolate each time step separately. This is performed by learning a function $\mathbf{f}(M, t) \rightarrow \mathbf{s}_t$, where $M$ is the initial stellar mass, $t$ is the time (age) to which it should be evolved, and $\mathbf{s}_t$ is a vector where each entry denotes a parameter of choice describing the star system at time $t$ (e.g., luminosity). This is accomplished in two ways, with a neural network and with hierarchical nearest-neighbor interpolation. Both approaches address the timescale variance problem with an age proxy. This means that instead of learning a function of $M$ and $t$ they learn approximations to two functions separately. One that maps $t$ to $p$, where $p$ is the age proxy and represents cumulative movement on the HR-diagram and another function $\mathbf{f}(M, p) \rightarrow \mathbf{s}_t$ \citep{maltsev_scalable_2024}. This very closely resembles the secondary EEPs as both the age proxy and the secondary EEPs are defined as movement along the HR-diagram. 

Unfortunately neither of these approaches can be directly applied to binary evolution. Both primary and secondary EEPs are not applicable to binary simulations since binary interactions alter stellar interiors in such ways that the standard single-star EEPs are no longer valid and cause aggressive changes in signal morphology. While an age proxy could in theory be designed for an approach like that of \cite{maltsev_scalable_2024} for binary simulations our early attempts did not show promising results, so we developed a different approach presented here.


\subsection{Relevant Problems in Signal Processing}

For the sake of completeness, we outline here the two most relevant formulations of the time-series generation problem found in the signal-processing literature. We discuss why neither can be actually applied to our problem of binary-evolution track interpolation.  The first problem formulation, which is very much prevalent in the signal-processing community, is that of \textit{time series forecasting}. It has been well studied due to the plethora of applications such as earthquake prediction, energy forecasting, climate forecasting, economic forecasting, and much more.

Traditionally these problems are solved without ML and instead involves using characteristics of historic data such as trend and seasonality of the signal to predict the next few timesteps of the signal. Classical approaches to this problem involve using either exponential or moving average smoothing to fit a function that models the function behavior in the near future. Key principles behind these approaches culminate in the Autoregressive Integrated Moving Average (ARIMA) \citep{kenett_autoregressiveintegrated_2014} model which is also extended to inject information regarding seasonality in Seasonal ARIMA (SARIMA). 

In the past decades, with the advent of big data, advances in hardware, and deep learning there has been growing interest in using Recurrent Neural Networks (RNNs) for time series forecasting. Most notably, Long Short Term Memory (LSTM) \citep{hochreiter_long_1997} neural networks have played a key role in advancing the field. More recently there has been growing interest in using principles behind the massively successful Transformers \citep{vaswani_attention_2023} for time series forecasting \citep{das_decoder-only_2024}. Unfortunately, these approaches are not directly applicable to our problem formulation given that we do not possess historic data of any signals. We also want to predict signals in their entirety instead of a few timesteps at a time. Finally, much of the literature surrounding this problem formulation assumes that the signals are uniformly sampled which is not the case for our problem.

The second problem formulation that we identified --- which is more applicable, but far less prevalent in the literature --- is that of \textit{time series generation}. Which, as the name implies, involves generating time series data which is similar to time series data in some dataset \citep{li_causal_2023, liao_conditional_2023, yuan_diffusion-ts_2024, kang_gratis_2020, madane_transformer-based_2022}. All of the works that we identified take a deep learning approach, obscuring the interpretability of their methods which may be undesirable for many scientific applications. Some key differences between our formulation and the ones in the literature are that our data are sampled on a grid and is therefore conditioned on specific initial conditions of the track. This has the ramification that these approaches do not assess the performance of their models on a testing set like we do, but rather assess the fidelity of the distribution of the generated time series to the distribution of the original time series. The problem formulation also lacks a defined similarity of signals which is given by the initial conditions in our formulation. This means we cannot condition signal generation on samples with the same characteristics as our target output. Thus, we cannot directly apply these techniques to solve our problem. 

\section{Proposed Method for Binary-Evolution Track Interpolation}

\subsection{Determination of Changepoints}

As mentioned earlier the problem we try to solve is to predict a track that would have been generated by a stellar evolution code like {\tt MESA} at a point $\mathbf{i}_* = (M_1^*, M_2^*, P^*)$ which does not exist in the given grid $G$. We provide such a prediction by utilizing the tracks corresponding to the neighboring grid points of the point $\mathbf{i}_*$. 
In developing such an interpolation method we face a number of challenges. Most importantly, we need to identify key points along the tracks where significant morphology changes occur without having clear astrophysical guidance, like in the case of single stars. Henceforth, we refer to these key sample points, as \textit{changepoints}. Moreover, the 
tracks are non-uniformly sampled and the number of sample points per track varies significantly across the grid. It should be noted here that the number of sample points per track is the same for all 10 parameters at a given grid point $\mathbf{i}_*$. To select the correct number of changepoints to represent a track we employ hyperparameter tuning which is described in Section \ref{section:4.6}.

\begin{figure}[h]
    \centering
    \includegraphics[width= 10cm]{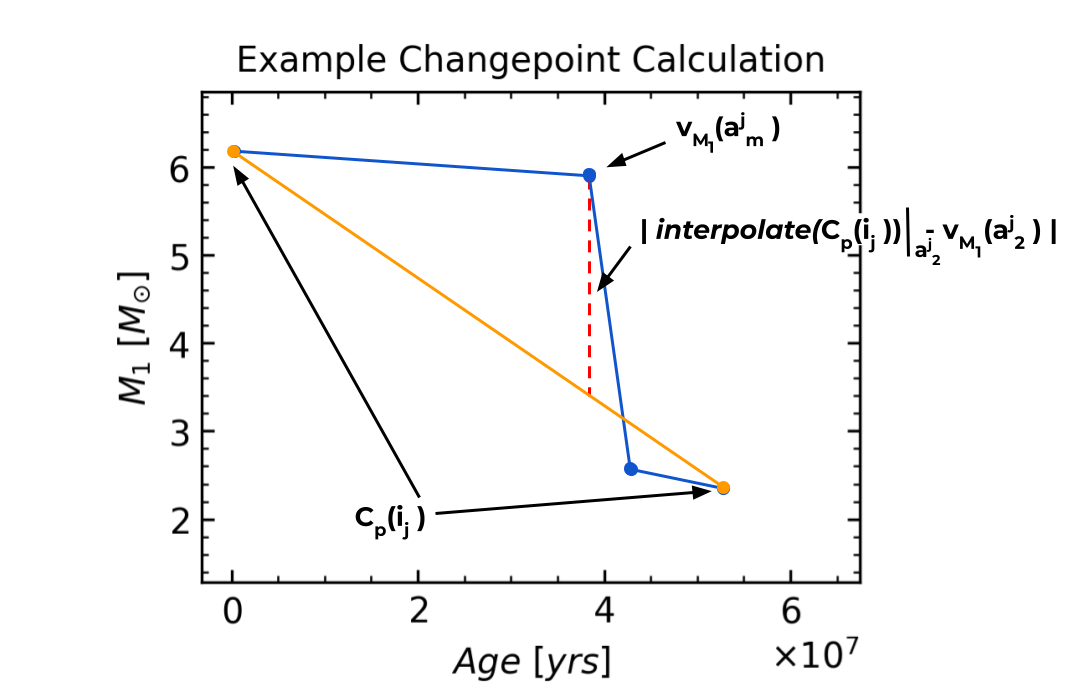}
    \caption{A visual illustration of how changepoints are computed. The blue signals are the original simulated tracks, the orange signal is a linear interpolation of the initial set of changepoints, and the red dashed line is the residual between the orange and blue signals.}
    \label{fig:changepoint_example}
\end{figure}

We propose the following algorithm based on \citet{900246} to define the changepoints for any given track. For an already simulated track $S_p(\mathbf{i}_j)$ in the grid as defined in Equation \ref{equation:simulation} and for a predetermined number of changepoints, $D$, we initialize the approximating track $C_p(\mathbf{i}_j)$ as

\begin{equation}
   C_p(\mathbf{i}_j) = \{ v_p({a_{1}^j}), v_p({a_{F(\mathbf{i}_j)}^j}) \}.
\end{equation}

\noindent Linear interpolation is performed using the set of changepoints in order to produce the set

\begin{equation}
   C'_p(\mathbf{i}_j) = \{ \mathrm{interpolate}(C_p(\mathbf{i}_j)) |_{a_n^j}, n = 1, \dots, F(\mathbf{i}_j) \}, 
\end{equation}

\noindent where the $\mathrm{interpolate}$ function performs linear interpolation between the set of changepoints as shown in Figure \ref{fig:changepoint_example}. Next, we find the time point $m$ which produces the highest approximation error, that is, 

\begin{equation}
    m^* = \underset{a_m^j}{\arg\max} | S_p(\mathbf{i}_j) - C_p'(\mathbf{i}_j)|, m = 1, \dots F(\mathbf{i}_j),
\end{equation}

\noindent
and the updated set of changepoints is defined as
\begin{equation}
    C_p(\mathbf{i}_j) = C_p(\mathbf{i}_j) \cup \{v_p(a_{m^*}^j)\}.
\end{equation}

\noindent This process is repeated until $D$ changepoints is reached.

\begin{figure}[h]
    \centering
    \includegraphics[width=\linewidth]{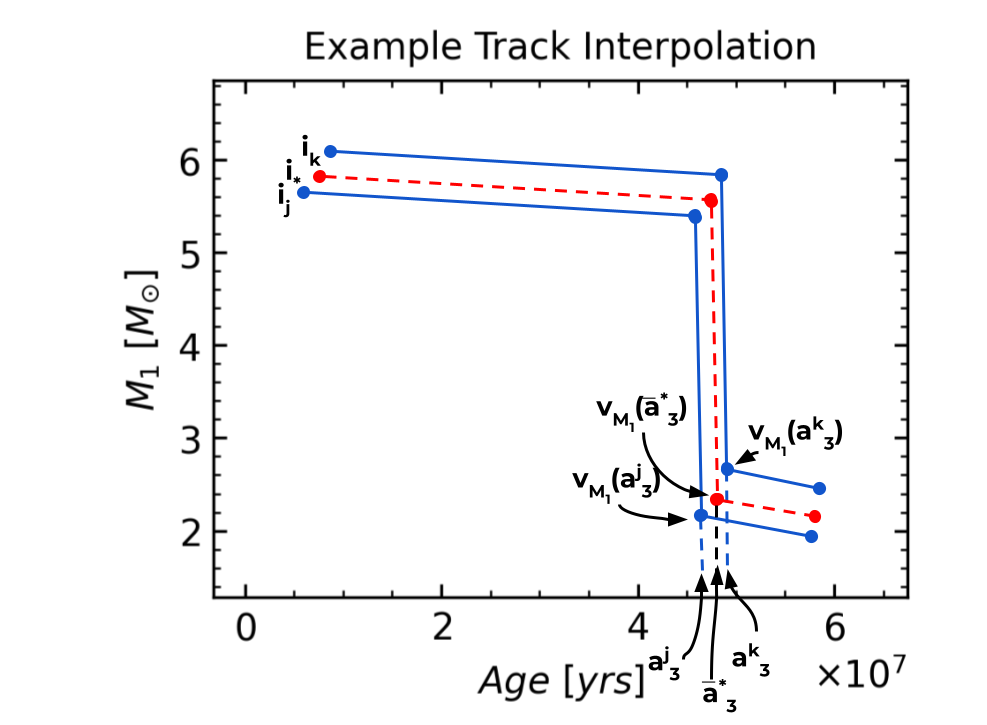}
    \caption{A visual illustration of how interpolation is carried out using the changepoint algorithm. The blue signals are the simulated tracks of the nearest neighbors of $\mathbf{i}_*$, shown in red. Equal weights ($w_1 = w_2 = 0.5)$ are used in this example.}
    \label{fig:changepoint_algorithm}
\end{figure}

Using this algorithm we effectively downsample the signals in our grid to contain only $D$ points from the given $F(\mathbf{i}_j)$ points of the track, where $D \leq F(\mathbf{i}_j)$. If $D > F(\mathbf{i}_j)$ we upsample the track such that $F(\mathbf{i}_j) \geq D$ before carrying out the changepoint algorithm. This process finds and preserves the most sizable/significant changes in the track, thus creating an alignment between tracks similar to \cite{dotter_mesa_2016}, and making effective interpolation possible. 

\subsection{$k$NN Interpolation}

Once we align our signals using the changepoint algorithm we can move on to actual track interpolation. Given the $k$ nearest neighbors (for barycentric interpolation which is discussed in Section \ref{section:4.3}, $k=4$) of some gridpoint $\mathbf{i}_*$ and their corresponding changepoints, we define the age points of our interpolated track as:

\begin{equation}
    \bar{a}_{i}^{*} = \sum_{j \in N(\mathbf{i_*})} w_j \cdot a_i^j
    \label{equation:8}
\end{equation}

\noindent where $N(\mathbf{i}_*)$ denotes the set of $k$ nearest neighbors of $\mathbf{i}_*$, $w_j$ are the weights that need to be determined as discussed later, with $\sum_{j \in N(\mathbf{i}_j)} w_j = 1$, and $a_i^j$ denotes the age of the $i$th changepoint for a track with initial conditions at $\mathbf{i}_j$. The value of the approximated parameter $p$ at $\bar{a}_j^* $ is determined as

\begin{equation}
    \hat{v}_p(\bar{a}_i^*) = \sum_{j \in N(\mathbf{i}_*)} w_j \cdot v_p(a_i^j).
    \label{equation:value}
\end{equation}

\noindent
The estimated track with initial conditions $\mathbf{i}_*$ is therefore given by 

\begin{equation}
    \hat{S}_p(\mathbf{i}_*) = \{\hat{v}_p(\bar{a}_1^*), \hat{v}_p(\bar{a}_2^*), \dots, \hat{v}_p(\bar{a}_D^*)\}.
    \label{equation:interpolated}
\end{equation}

\noindent
A visual illustration of the algorithm is given in Figure ~ \ref{fig:changepoint_algorithm}. 

\subsection{Determination of Interpolation Weights}
\label{section:4.3}

We now turn to the problem of finding an appropriate set of weights to be used in Equations \ref{equation:8} and \ref{equation:value}, based on the Euclidean distance, $d$, of $\mathbf{i_*}$ to its closest neighbors \footnote{We define $d$ as the Euclidean distance between initial conditions in $\log_{10}$ space, that is $\log_{10}(M_1^j)$, $\log_{10}(M_2^j)$, and $\log_{10}(P)$.}. We considered calculating the weights based on various functional forms of $d$, for example, $w_j = d_j^{-k}$, for $k = 1, 2, 3$. We also considered barycentric weights. They are defined as barycentric coordinates \citep{berrut_barycentric_2004} of a point with respect to a shape and thus require a tessellation of $G$. We use a Delaunay triangulation \citep{barber_quickhull_1996} over the set of initial conditions $G$ to construct a convex hull where each vertex is a signal's initial conditions. To carry out interpolation of a point $\mathbf{i}_*$ with some initial conditions --- $M_1^*$, $M_2^*$, and $P^*$ --- we find which tetrahedral $T$ the point falls into. We then compute the barycentric coordinates of $\mathbf{i}_*$ with respect to $T$. As mentioned before, these coordinates are then used to define the weights. We find that this latter choice performs much better than the simple functional forms of $d$, so we adopt it moving forward. One drawback to this weighting scheme is that if $\mathbf{i}_*$ is outside the described convex hull, we have no means of performing interpolation so we resort to using the nearest neighbor as the approximation.

\subsection{Signal Classification}

\begin{figure}
    \centering
    \includegraphics[width=\linewidth]{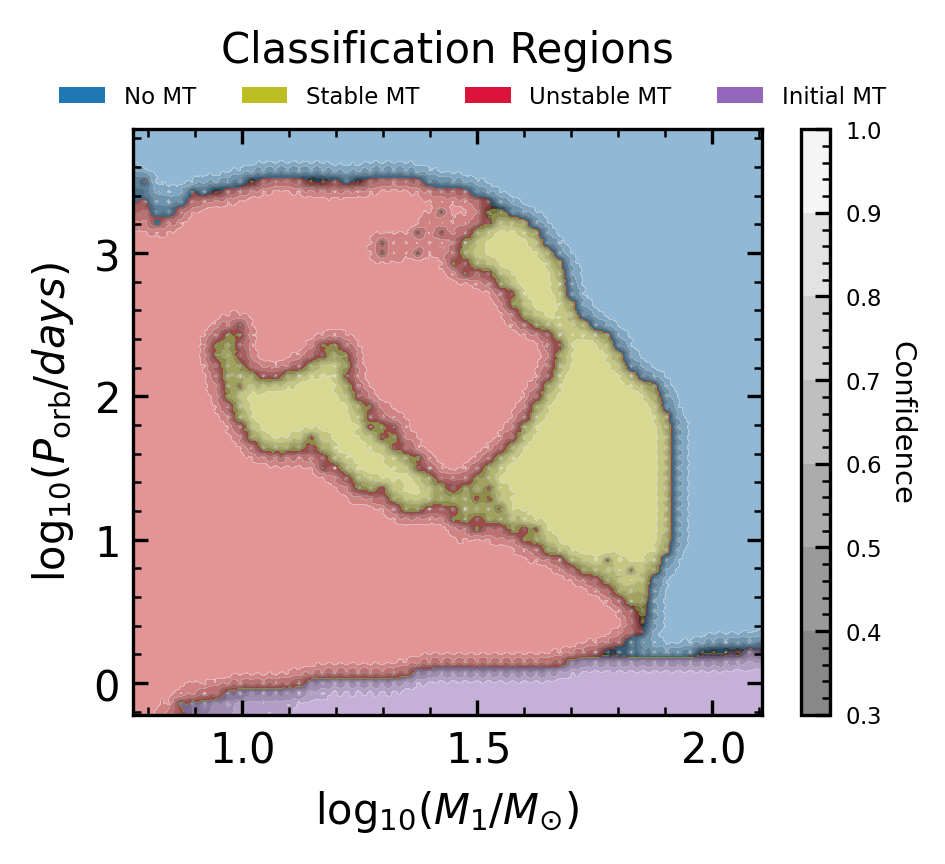}
    \caption{Example case of a two-dimensional slice of one of the binary grids. Classes are selected based on the type of mass transfer in the binary track. Classification regions are color-coded while classification confidence is indicated by an overlaid layer which ranges from transparent (complete confidence) to opaque (no confidence). It is evident that classification confidence, and thereby accuracy, decreases close to class boundaries, as one would expect.}
    \label{fig:grid_slice}
\end{figure}

The morphology of tracks in any random neighborhood of $G$ is not guaranteed to be similar. To address this we employ the classification scheme used for initial/final interpolation in \cite{fragos_posydon_2023}. This scheme classifies tracks by the mass transfer type exhibited through $k$-nearest neighbors, where $k$ is set using Monte Carlo Cross Validation (MCCV) and the weight to each neighbor is the inverse distance between $\mathbf{i}_*$ and its $k$ neighbor squared ($d^{-2}$)\citep{fragos_posydon_2023}. Classification regions and confidence for this classifier can be seen in Figure \ref{fig:grid_slice}. We observe that classification confidence is high in regions of class homogeneity which make up large parts of the two-dimensional slice. As one might expect, classification confidence decreases around class boundaries due to the rapid changes in grid-point class. This is what active learning methods can improve (e.g., \cite{rocha_active_2022}).

 In our experiments the mass-transfer classes defined mostly dictate the morphology of the signals, at least enough for this classification scheme to be used effectively in practice. In our experiments this classification approach outperforms $k$-medoids \citep{arora_analysis_2016} clustering approaches. Since signals close in the grid within a class most likely have similar morphology we can now use a $k$-nearest neighbor approach to produce an approximation for any signal with initial conditions within the range of initial conditions of $G$. Note that when using Barycentric weighting the Delaunay triangulations must be computed for each class separately, and interpolation is carried out by using the corresponding convex hull of the predicted class label of an initial condition $\mathbf{i}_*$.

\subsection{Special Treatment of \texorpdfstring{$\log_{10}(\dot{M}_\mathrm{transfer})$}{Mass-Transfer Rate}}

The proposed method works well for almost all signal morphologies as will be described later. However, often the tracks associated with the $\log_{10}(\dot{M}_{\mathrm{transfer}})$ parameter exhibit a morphology that requires special treatment. If there is no mass transfer the shape of the $\log_{10}(\dot{M}_{\mathrm{transfer}})$ signal presents no issue for the interpolation problem. Mass transfer in our simulations often happens for a very short period of time (relative to the entire duration of the signal). This manifests itself as signals exhibiting one or more spikes resembling a dirac delta function (although when looking closely there are important differences). The morphology of the signal outside of this spike is constant which is straightforward to interpolate. In this case, we segment each of these signals into regions for which  $v_p(a^j_n) > \tau$, and regions for which $v_p(a^j_n) < \tau$, for some threshold $\tau$ (chosen to be equal to $10^{-99} M_{\odot} / yr$). The changepoint algorithm described earlier is then only applied within the former segment of the signal, when mass-transfer is active.

\subsection{Hyper-parameter Tuning}
\label{section:4.6}

\begin{figure*}
    \centering
    \includegraphics[width = \textwidth]{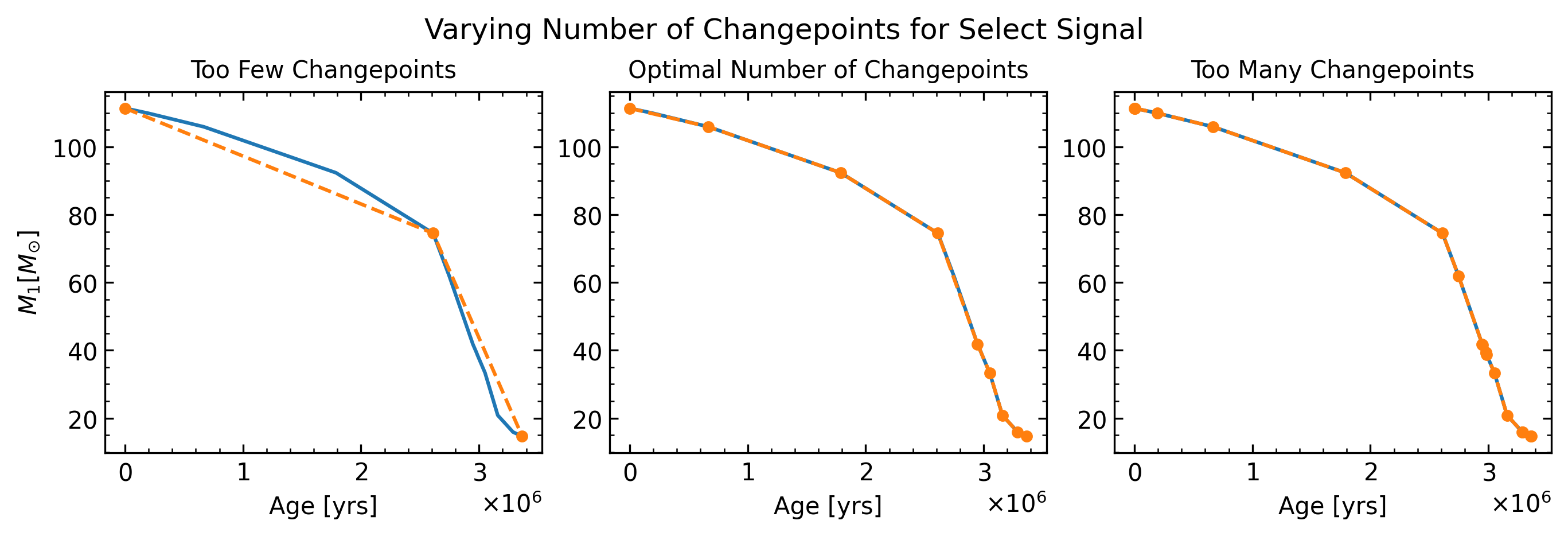}
    \caption{Effect of the number of changepoints on the approximation error. The blue solid curve represents the same original signal in all plots and the orange dotted curve represents the approximating signal which varies linearly between the changepoints shown as orange dots. The number of changepoints used is equal to 3, 9, and 18 from left to right. Clearly the left panel has too few changepoints to capture the morphology of the signal. Meanwhile the panel on the right captures the morphology quite well but as the number of changepoints is increased, the placement of the points becomes sensitive to small perturbations (noise) in the signal making an alignment between signals with similar morphology harder. Finally, the panel in the middle has an adequate number of changepoints to capture the morphology of the signal while also maintaining the alignment with signals that exhibit similar morphology.}
    \label{fig:figure_2}
\end{figure*}

For the proposed method we have two hyper-parameters to tune, namely $D$, the number of changepoints and $k$, the number of neighbors. For barycentric interpolation $k$ is fixed and equal to four; thus, we are left with optimizing for $D$. The main consideration concerning the value of $D$ is that if it is too small, then the morphology of the signal is not adequately captured. On the other hand, if it is too large, although the original signal is approximated quite accurately, the placement of the changepoints follows all the small changes in the value of the signal and therefore increases the chance of misalignment among the change points in neighboring tracks. 
\startlongtable
\begin{deluxetable*}{l | c c c c c c c c c c}
\tabletypesize{\scriptsize}
\tablewidth{0pt}
\setlength{\tabcolsep}{1.2\tabcolsep}  

\tablecaption{Optimal value of $D$ by class for each parameter in the HMS-HMS grid. The first column indicates the class while the remaining columns indicate the parameter. Each cell of the table indicates the number of changepoints.\label{table:hyperparameters}}
\tablehead{
    \colhead{$Class$} &
	\colhead{$M_1$} &
	\colhead{$M_2$} &
	\colhead{$P_{orb}$} &
	\colhead{$\log_{10}(\dot{M}_{transfer})$} &
 	\colhead{$\log_{10}(T_\mathrm{eff, 1})$} &
	\colhead{$\log_{10}(L_1)$} &
    \colhead{$\log_{10}(R_1)$} &
    \colhead{$\log_{10}(T_\mathrm{eff, 2})$} &
	\colhead{$\log_{10}(L_2)$} &
    \colhead{$\log_{10}(R_2)$}
    }
\startdata
stable MT & 16 & 16 & 32 & 16 & 32 & 64 & 32 & 64 & 64 & 64 \\
unstable MT & 16 & 32 & 16 & 32 & 32 & 32 & 16 & 32 & 32 & 32 \\
no MT & 16 & 16 & 16 & 4 & 16 & 16 & 16 & 32 & 32 & 32 \\ 
\enddata
\end{deluxetable*}

The effect of the number of changepoints $D$ on the approximation error can be seen in Figure \ref{fig:figure_2}, in which an original signal is shown along with 3 approximations of it. In the the right-most panel changepoints are placed on top of one another at points where the change is imperceptibly small and can be due to numerical noise or other insignificant changes. When applying the changepoint algorithm to a neighboring track the placement of changepoints may correspond to noise at a different part of the track. Thus, the changepoints among neighboring tracks may no longer align. Therefore, it is important to constrict the number of changepoints to only capture the most significant morphological changes of each track. Since the morphology of the tracks in each class may differ, we utilize a different number of changepoints per class.

To do this we used a validation set to find the $D$ that produces the smallest error. Note that because many parameters can describe a simulation, we need to optimize $D$ for each of the 10 parameters of the signal independently and also perform the interpolation independently. We show the number of changepoints used by class for each parameter in Table \ref{table:hyperparameters}. We used the validation set to assess the performance of using 4, 8, 16, 32, and 64 changepoints and chose the best performing combination of hyper-parameters to use in our final algorithm.

\subsection{Applying Physical Constraints}

Given that our goal is to approximate physical processes, it is important to ensure that  our approximations to follow known laws of physics. To do this we carry out a post processing step that enforces relationships between the parameters we are interpolating. This step applies the same constraints to our interpolated quantities as in \cite{fragos_posydon_2023}. For example, in connection to the ten parameters considered here, we enforce the Stefan-Boltzmann law for both stars which connects the effective temperature $T_\mathrm{eff}$, luminosity $L$, radius $R$, and is given by

\begin{equation}
    T_\mathrm{eff} = (L / 4 \pi R^2 \sigma_\mathrm{SB})^{1/4}.
    \label{equation:stefan-boltzmann}
\end{equation}

\begin{figure*}
    \centering
    \includegraphics[width=\textwidth]{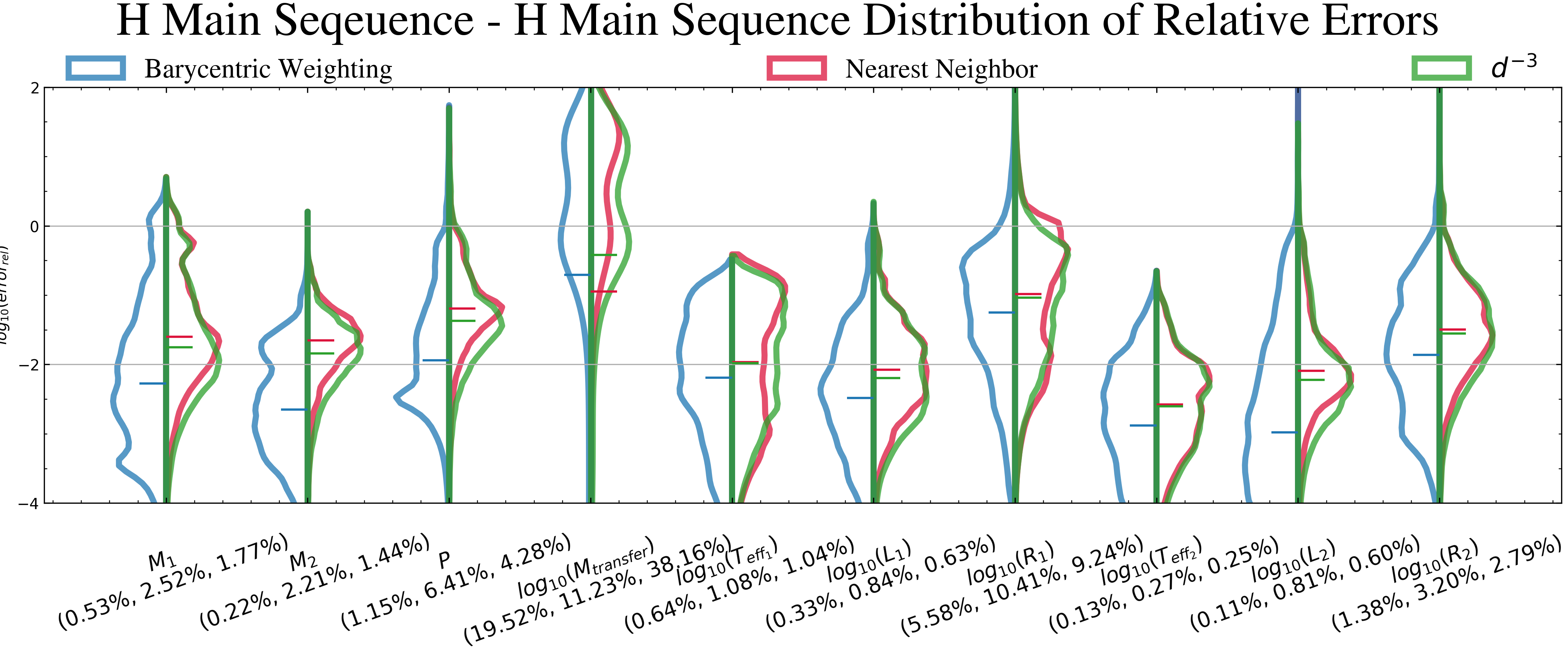}
    \caption{A comparison of three different nearest neighbor methods. For each parameter three different histograms of errors are shown where errors are calculated using Equation \ref{error:rel}. The blue and green histograms correspond to the proposed method with barycentric and inverse distance cubed weights, and the red corresponds to the nearest neighbor. The implementation using the inverse distance cubed as weights uses $k = 4$. Median values are marked within each histogram and are also indicated on the $x$-axis in the following order, barycentric weighting (blue), nearest neighbor (red), and $d^{-3}$ (green).}
    \label{fig:hms-hms-discrete}
\end{figure*}

To enforce the equation we pick the two best performing parameters (as given by the validation set) and compute the third parameter using Equation \ref{equation:stefan-boltzmann}. It should be noted that during the HMS-HMS phase this constraint is applied to both stars in the system. More broadly, other such constraints (conservation laws, other physical relations between parameters, physical expectations for monotonicity) can be imposed in a post-processing step.

\section{Evaluation}

\subsection{Evaluation Methods}

 In \cite{dotter_mesa_2016} evaluation is carried out visually while in \cite{maltsev_scalable_2024} evaluation is performed by comparing L1, L2, and L$\infty$ errors. We argue that while these evaluation metrics are sufficient for single star track interpolation, binary star track interpolation requires special considerations due to the extreme morphologies of the signals during binary interactions. 

\subsection{Challenges of Evaluation}

A typical evolution of $M_1$ can be seen in panel d) of Figure \ref{fig:problem_description}. The signals are often characterized by a large negative slope for the majority of the evolution, which is followed by a relatively small period of extremely small negative slope visually appearing as a discontinuity caused by mass transfer. If the approximation predicts the onset of mass transfer too late or too early extremely high errors are exhibited. Furthermore, MESA samples at a higher rate during parts of the evolution where the system is subject to relatively large changes. The consequence of this is that there are disproportionately many time steps where the system is exhibiting mass transfer--which is the region of the track most difficult to approximate. Thus, the placement of timesteps are biased toward producing high L1, L2, or L$\infty$ errors. However, high accuracy in these regions may not be important, rather a small difference in morphology between predicted and ground truth tracks is paramount. This poses a challenge in evaluation and hyperparameter tuning.

\subsection{Our Evaluation Method}

\begin{figure*}
      \centering
      \includegraphics[width=0.3\textwidth]{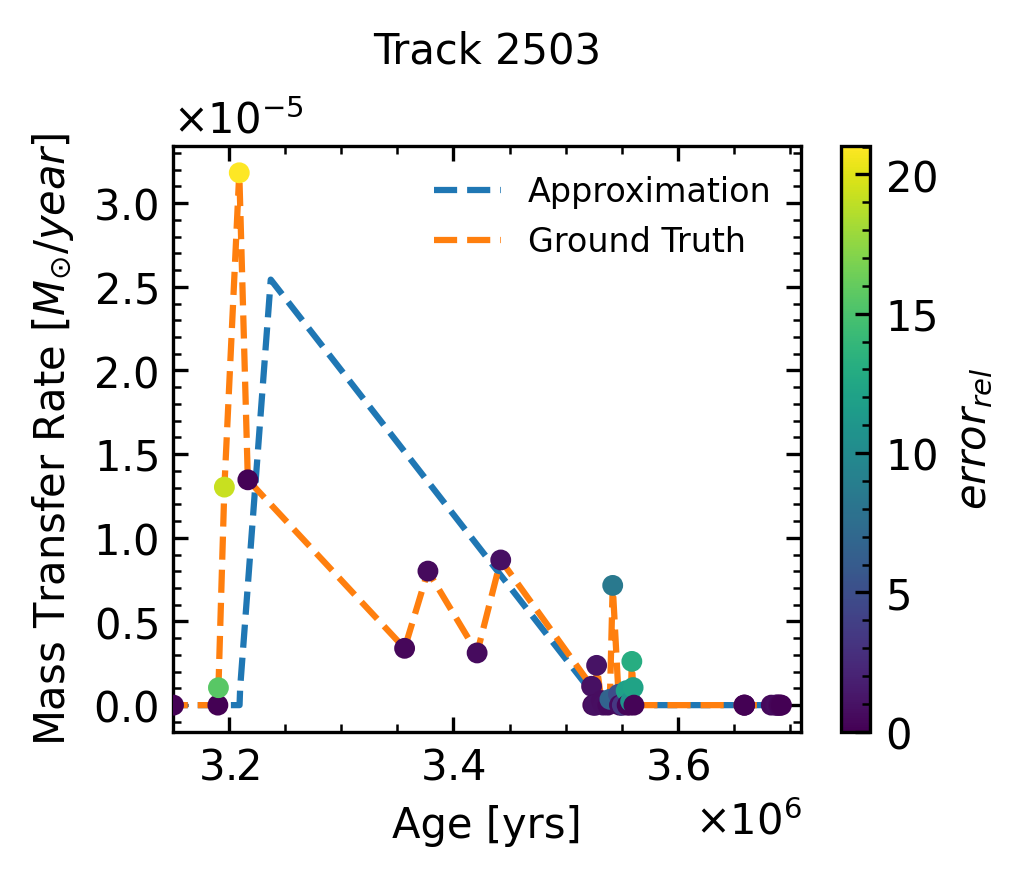}
      \hfill
      \includegraphics[width=0.3\textwidth]{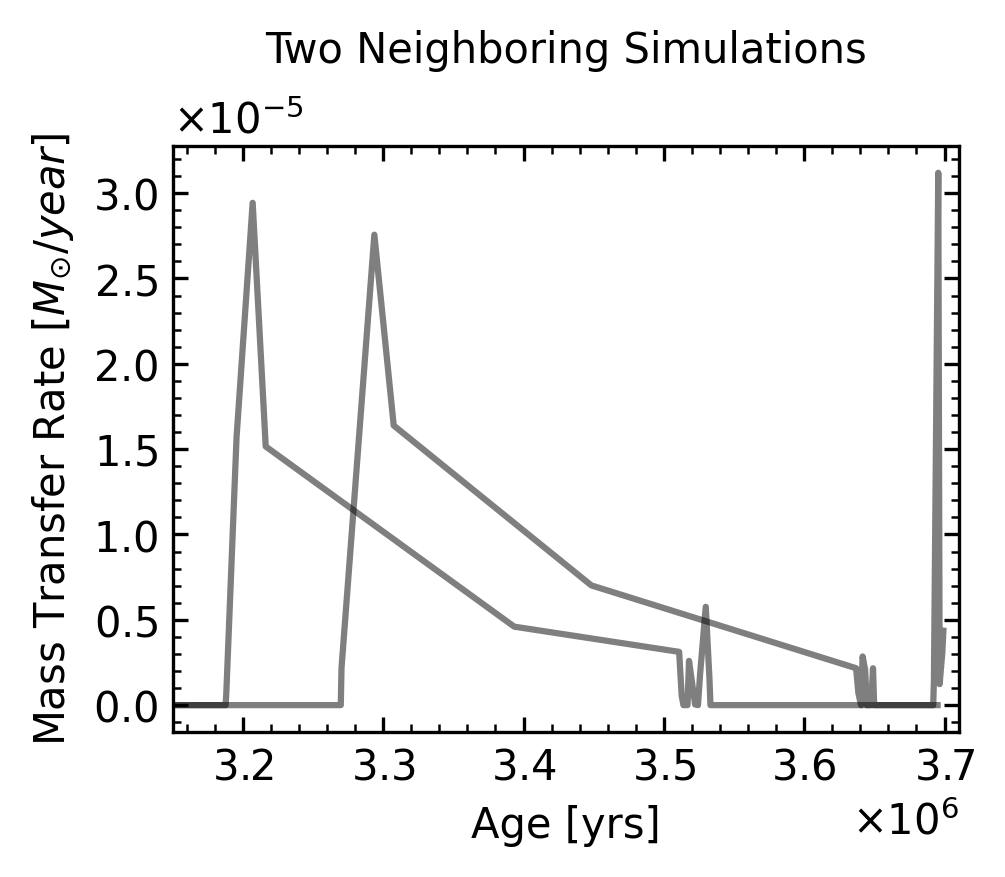}
      \hfill
      \includegraphics[width=0.3\textwidth]{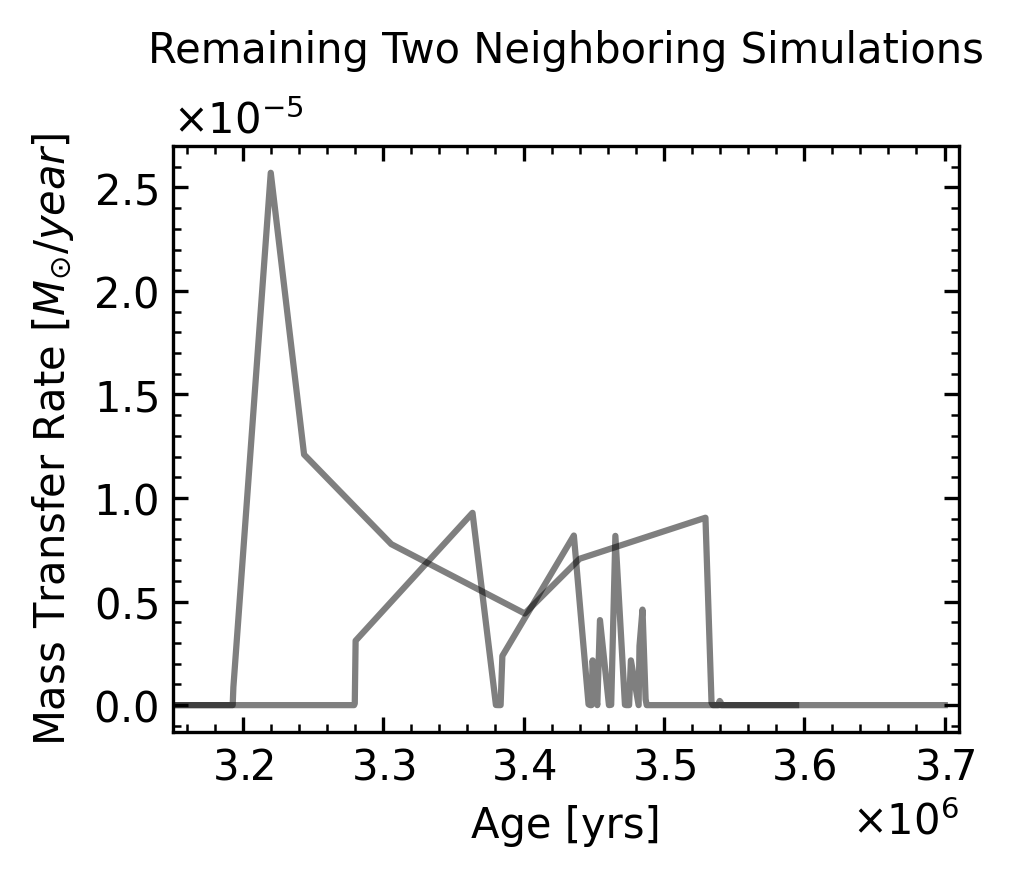} \hfill
    \caption{An example of a $\log_{10}(\dot{M}_{\mathrm{transfer}})$ signal that is particularly difficult to interpolate due to non uniform morphology of nearest neighbors. In the leftmost panel the dashed orange and blue lines correspond to the ground truth and interpolated tracks respectively, and the colored markers on the ground truth represent the ground truth timesteps used for evaluation while the color corresponds to the severity of error. The middle and ri show translucent black lines which correspond to the four neighbors used for interpolation.}
    \label{fig:mtransfer}
\end{figure*}

\begin{figure*}[ht]
    \centering
    \includegraphics[width=\linewidth]{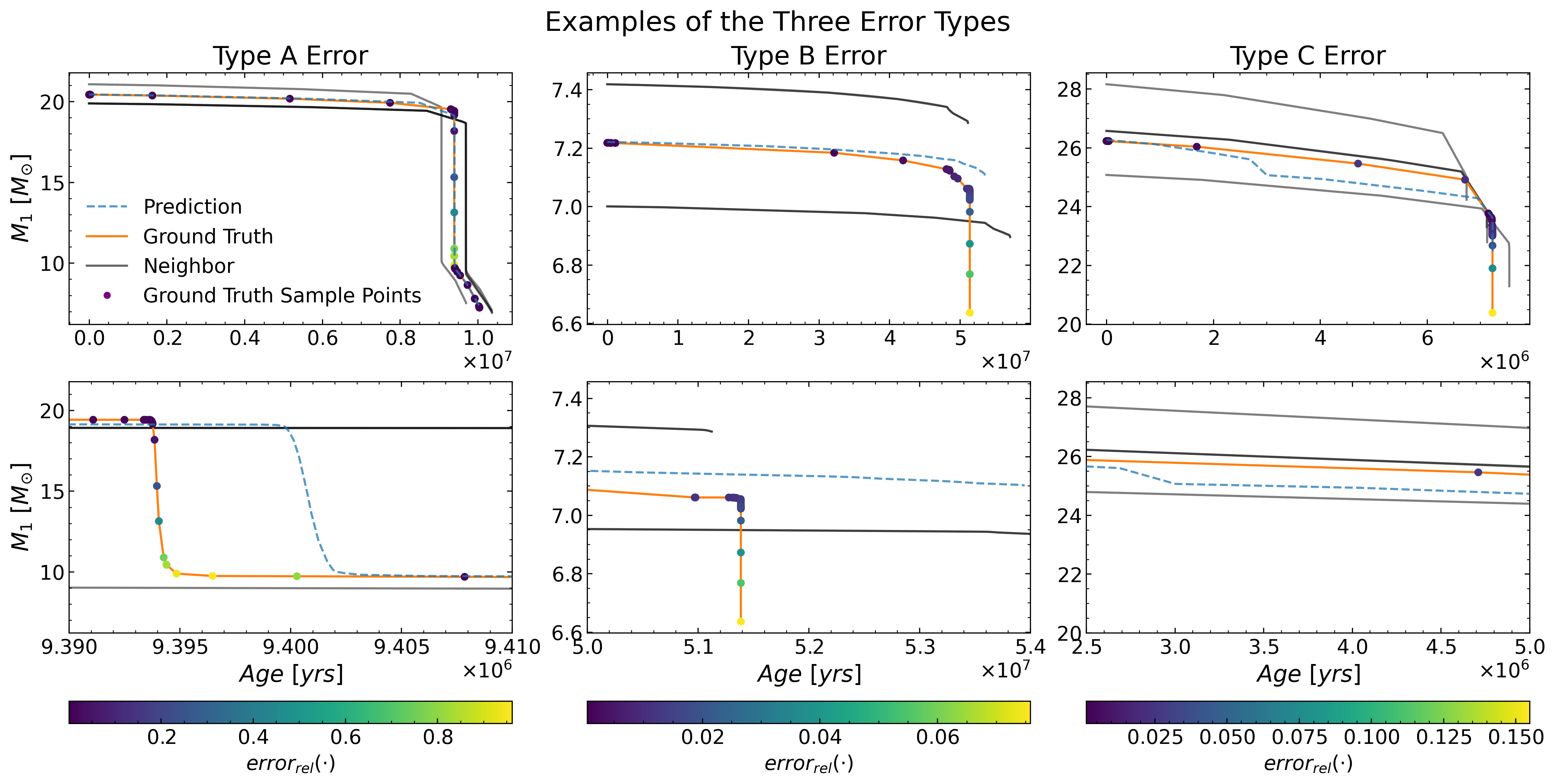}
    \caption{A summary of the different error types we have identified. Each column corresponds to an example different error type. The top row shows the complete tracks while the bottom row shows a zoomed in view of the tracks. The interpolated track is shown in blue, the ground truth is shown in orange, and the nearest neighbors used are shown in black. The colored dots show the error produced at each sampled time in the ground truth.}
    \label{fig:error_types}
\end{figure*}

\begin{figure*}[htp]
    \centering
    \includegraphics[width=\textwidth]{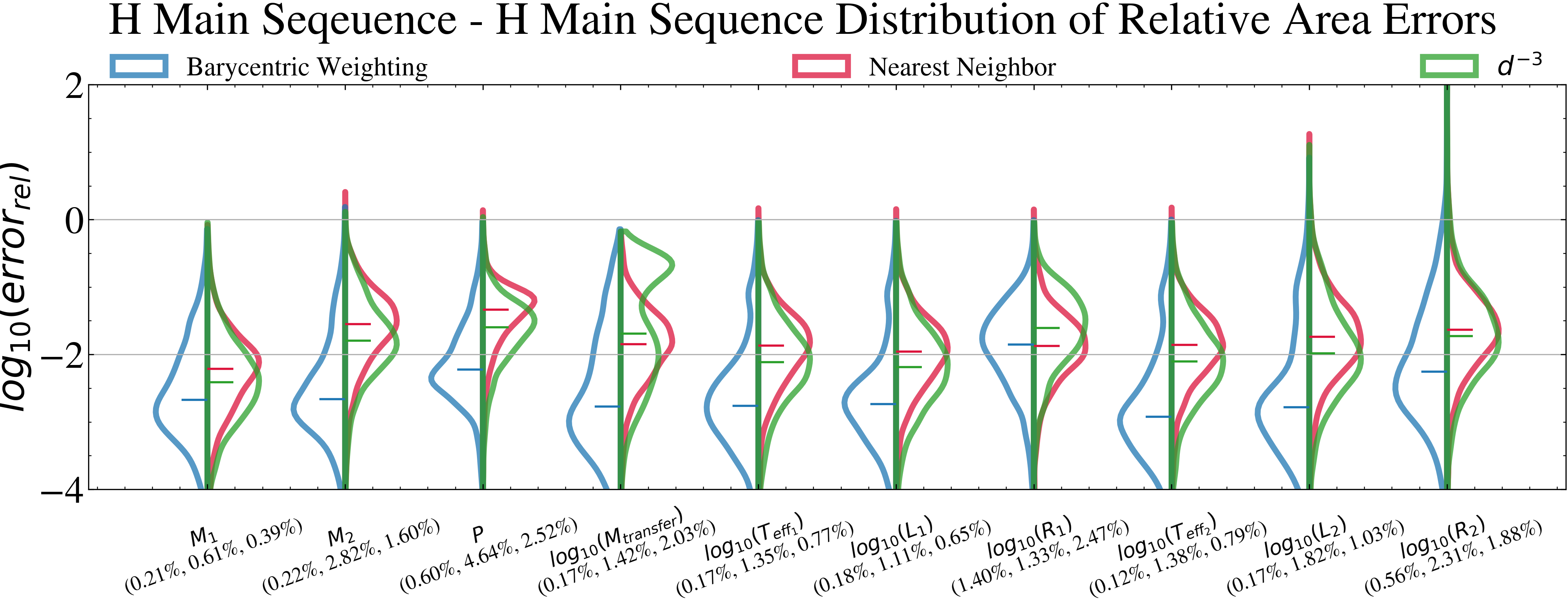}
    \caption{A comparison of three different nearest neighbor methods. For each parameter three different histograms of errors are shown where errors are calculated using Equation \ref{error:area}. The blue and green histograms correspond to the proposed method with barycentric and inverse distance cubed weights, and the red corresponds to the nearest neighbor. The implementation using the inverse distance cubed as weights uses $k = 4$. Median values are marked within each distribution and are also indicated on the $x$-axis in the following order, barycentric weighting (blue), nearest neighbor (red), and $d^{-3}$ (green).}
    \label{fig:hms-hms-continuous}
\end{figure*}

To address this we propose a suite of evaluation methods that offer a more comprehensive view of the algorithm's performance. Evaluation is carried out using a set of simulations generated with initial conditions randomly sampled within the range of $G$. We then compute the relative error between our approximation and its ground truth for every time step, $a_n^*$, for each parameter that we are interpolating independently. To do this, we first define a continuous version of our discrete signal $S_p(\mathbf{i}_*)$ as

\begin{equation} \label{error:continuous_signal}
    S^c_p(\mathbf{i}_*) = \mathrm{interpolate}(S_p(\mathbf{i}_*)).
\end{equation}

\noindent Therefore, the relative error for parameter $p$ at time point $n$ of $\hat{S}_p(\mathbf{i}_*)$ is defined as 

\begin{equation} \label{error:rel}
    \mathrm{error_{rel}}(\mathbf{i}_*, p, a_n^*) = \mid \frac{\hat{S}^c_p(\mathbf{i}_*)\mid_{a^*_n} - v_p(a^*_n)}{v_p(a^*_n)} \mid.
\end{equation}

\noindent The timestep error histogram of each parameter for the HMS-HMS phase of the evolution can be seen in Figure \ref{fig:hms-hms-discrete}. It is desirable that the proposed method,

\begin{enumerate}
    \item Produces a median error below $10\%$
    \item Outperforms the class-wise nearest neighbor method
\end{enumerate}

As is evident, our method meets both criteria for nine out of the ten parameters. Medians of the error distribution (shown in blue) in Figures \ref{fig:hms-hms-discrete} that are below $10\%$ are considerably lower than the medians of the class-wise nearest neighbor (shown in red). Furthermore, we show our proposed method with a different weighting scheme (shown in green). However while the medians across all parameters are quite low, all distributions exhibit outlier errors which are considerably high which led us to analyze the cause of these outlier errors. 

Even with our special treatment, interpolation of the $\log_{10}(\dot{M}_{\mathrm{transfer}})$ parameter presents a particular challenge for our method which will need to be further addressed in future work. Signal morphology exhibited in this parameter is dependent on class, while the interpolation for certain classes is straightforward, for the majority of cases the morphology resembles that of a dirac delta function. Interpolation for signals exhibiting this morphology is very difficult because a slight offset between the position of the dirac delta like spike between the interpolated and ground truth signals lead to large errors. Furthermore, we find many instances where neighbors do not exhibit similar morphology. This can be seen in Figure \ref{fig:mtransfer} which illustrates the difficulty of interpolating this parameter. In the figure three out of the four neighbors exhibit the same morphology, however, the third neighbor does not possess a spike as the others. The signals exhibit high frequencies which also contribute to difficulty in interpolation.

Through analysis of high errors we were able to determine the following three types of high errors which are illustrated in Figure \ref{fig:error_types}. 

\begin{enumerate}
    \item \textbf{Type A Error:} As mentioned before, a common morphology of our signals is characterized by large changes in signal value. An example of such a signal is shown in Figure \ref{fig:error_types} (leftmost column). The top panel shows the predicted track faithfully following the ground truth track. However, large errors occur around the times of the large change (drop) in signal value. This becomes apparent when zooming into that part of the time axis. The predicted track has the same morphology as the ground truth track, but it predicts that the signal value change occurs $\simeq\,10,000$ years later than in the ground truth track. This is of little concern from an astrophysical point of view. However, the errors in this short time segment determined the maximum value of the relative error. This was the most prevalent cause of high errors in our analysis, accounting for $\simeq\,70\%$ of high errors.
    \item \textbf{Type B Error:} This type of error occurs when the nearest neighbors used to produce the estimated track belong to different classes, due to misclassifications. An example of this type of error is shown in the middle column of Figure \ref{fig:error_types}. We see in the top figure that the two neighbors to the predicted track have a different shape and duration than the actual track to be predicted. As a result the predicted track incurs a large error, as also shown in greater detail in the bottom of this Figure. This issue is also relevant to Figure 7. 
    \item \textbf{Type C Error:} As mentioned before, signals require a specific number of changepoints to effectively create an alignment. Too few changepoints fail to adequately capture signal morphology while too many changepoints result in changepoints no longer corresponding to meaningful changes in signal value causing a change in morphology of the predicted track from the morphology of the nearest neighbors. In Figure \ref{fig:error_types} it is apparent that the approximation produced does not maintain the morphology of any of its neighbors and therefore causes the signal to deviate significantly from the ground truth signal. Fortunately, this type of error is the least prevalent and accounts for only $\approx 5\%$ of the worrisome errors.
\end{enumerate}

\begin{figure}
    \centering
    \includegraphics[width = \linewidth]{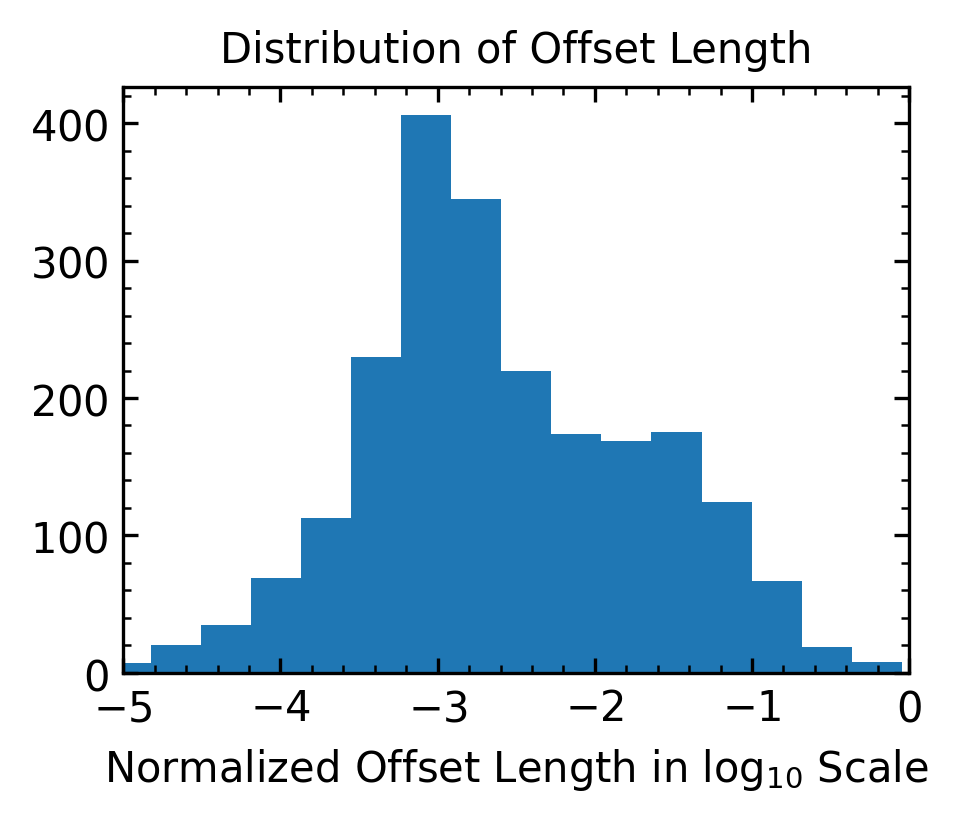}
    \caption{A histogram of the normalized age offsets of the large drops in the $M_1$ signal value between the predicted and ground truth tracks divided by the ground truth final age. 
    }
    \label{fig:offset_distribution}
\end{figure}

\noindent To better quantify the first type of error we propose computing the relative area between the ground truth and the predicted tracks, that is,  

\begin{equation} \label{error:area}
    \mathrm{error_{area}}(p, \mathbf{i}_*) = \mid \frac{\int \hat{S}^c_p(\mathbf{i}_*) \,dx - \int S^c_p(\mathbf{i}_*) \,dx }{\int S^c_p(\mathbf{i}_*) \,dx} \mid,
\end{equation}

\noindent where $x$ is the temporal axis of the evolution. 
The distribution of area errors for the 10 parameters under consideration can be seen in Figure \ref{fig:hms-hms-continuous}. Similarly to Figure \ref{fig:hms-hms-discrete} using the area error metric maintains medians below $10\%$ and outperforms the class-wise nearest neighbor interpolation when analyzing area errors for nine out of the ten parameters. To more comprehensively evaluate type A errors we also plot the histogram of the normalized (fractional/relative) age offsets of the large drops in the $M_1$ signal value between the predicted and ground truth tracks in Figure \ref{fig:offset_distribution}. The histogram peaks at around $10^{-3}$, meaning that the age offset between large drops in the $M_1$ signal value will be around $0.1\%$ of the simulation's final age.

\begin{figure*}
      \centering
      \includegraphics[width=0.33\textwidth]{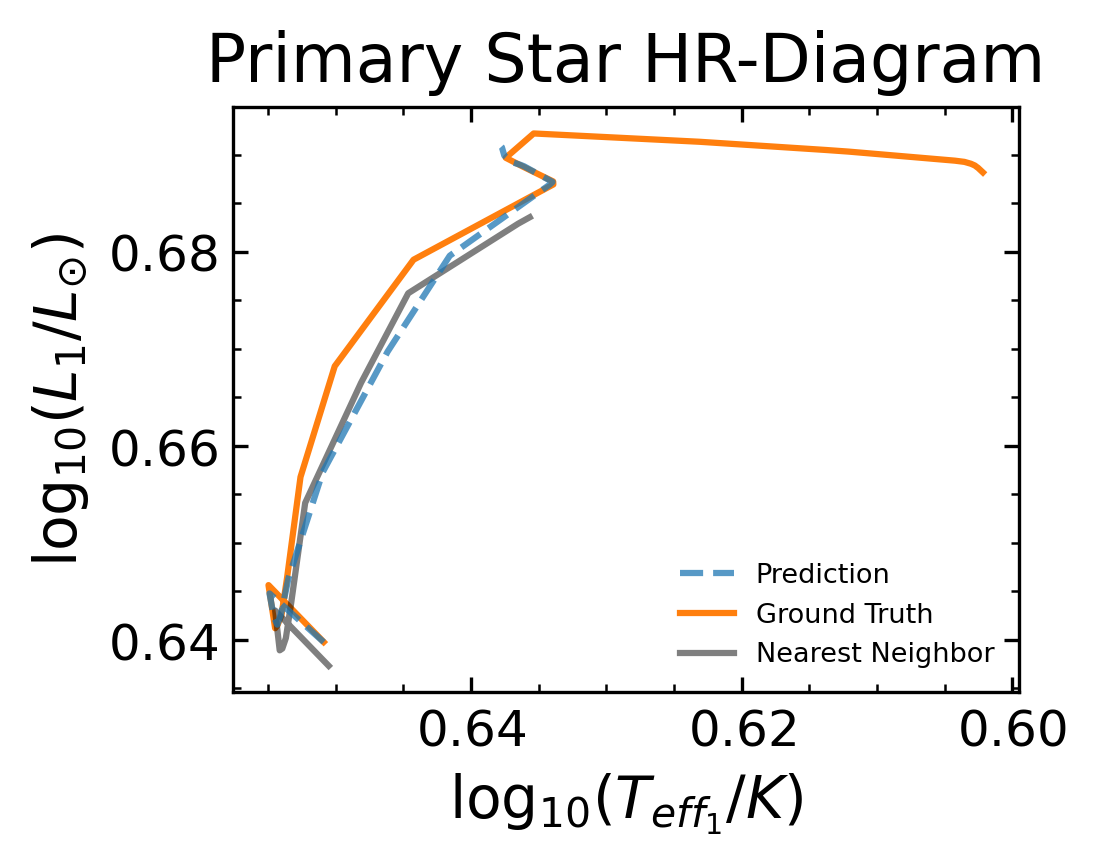}\hfill
      \includegraphics[width=0.33\textwidth]{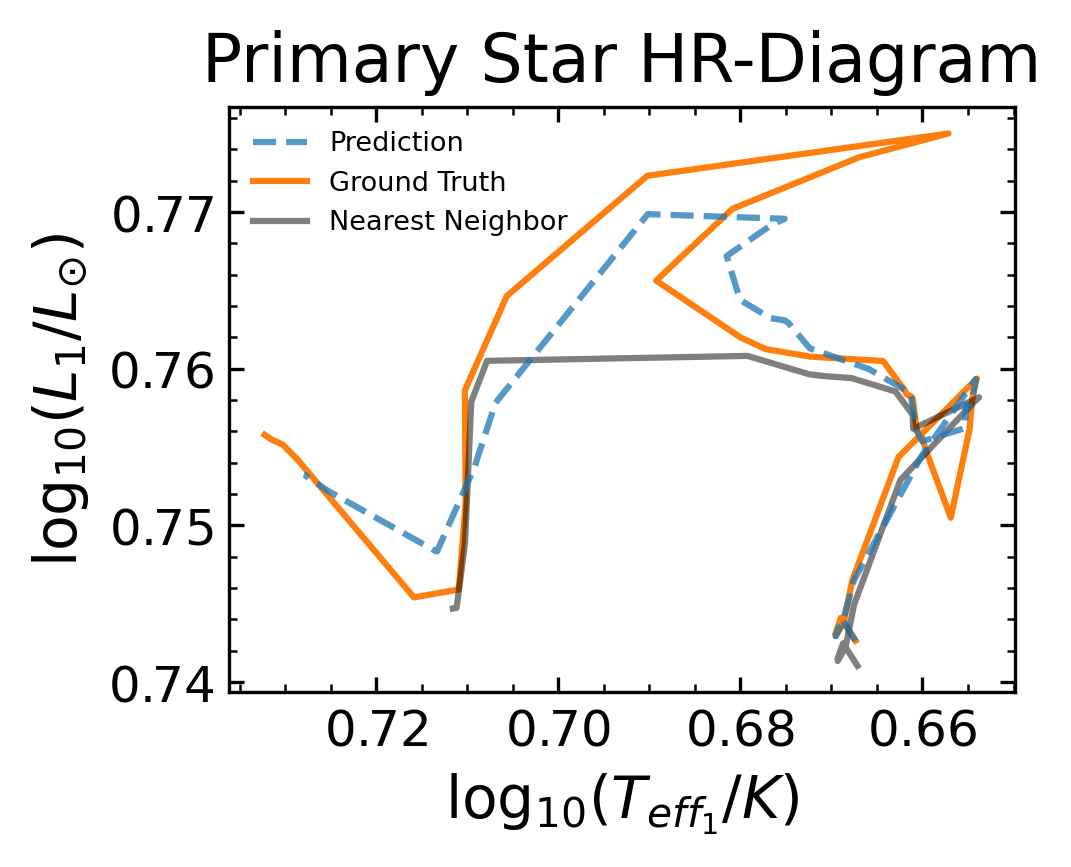}\hfill
      \includegraphics[width=0.33\textwidth]{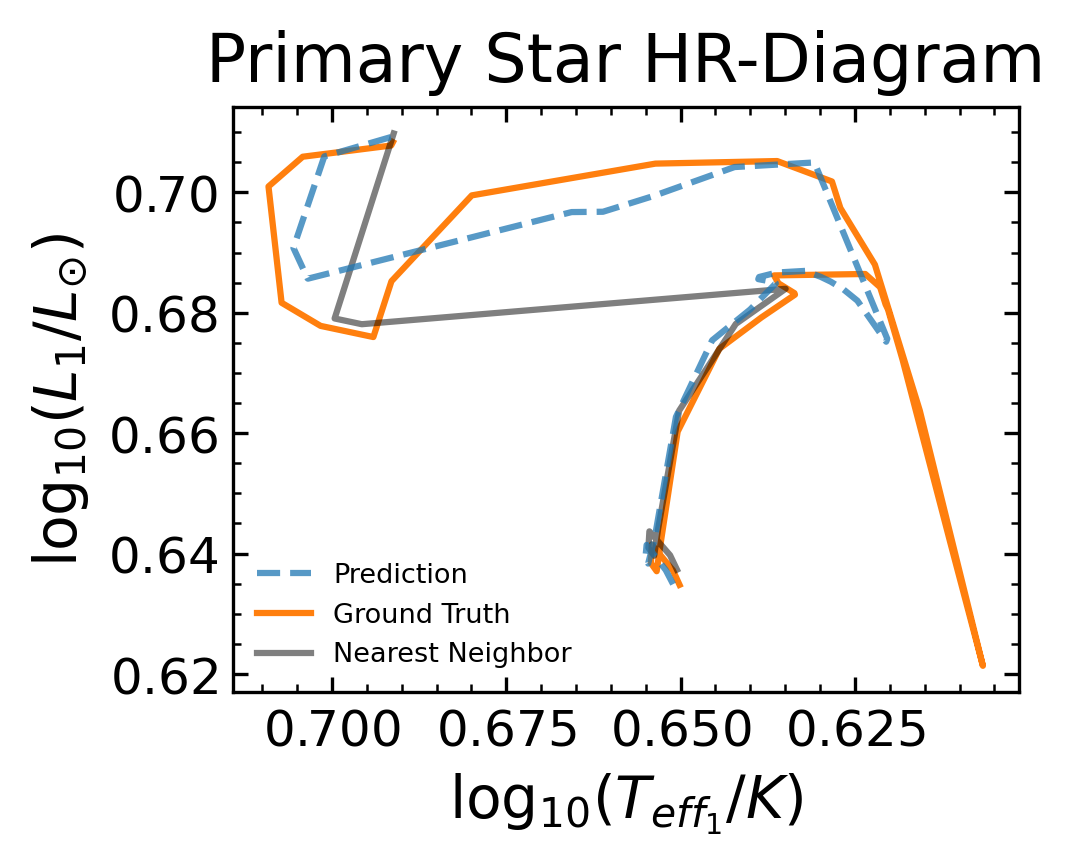}
    \caption{Three examples of HR-Diagrams of the primary star of each system. The leftmost panel corresponds to low error, the middle panel corresponds to medium error, and the rightmost panel corresponds to high error. The ground truth is given in blue while the approximation produced with our proposed method is given in orange. The $x$-axis is inverted and shows the $\log_{10}$ effective temperature while the $y$-axis shows the $\log_{10}$ luminosity.}
    \label{fig:hr_diagrams}
\end{figure*}

The second cause is errors in the classification step is largely unavoidable, and third cause is an inappropriate use of $D$, the number of changepoints used to represent the signal. This can be alleviated by more sophisticated classification methods that more strongly consider the morphology of the signals and hyper-parameter tuning.

In conjunction with quantitative evaluation methods we also turn to qualitative evaluation methods. To do this we picked predicted tracks with high, medium, and low errors and then plotted their HR-diagrams which can be seen in Figure \ref{fig:hr_diagrams}. Both HR-diagrams given by our proposed method corresponding to low and medium error faithfully follow the ground truth while also starting and finishing at roughly the same place. However, the HR-diagram corresponding to high errors shows a significant deviation between our the predicted track by our proposed method and the ground truth while ultimately beginning and ending at roughly the same place in the diagram. This shows that despite interpolating parameters independently we are still able to maintain the expected relationships between parameters as defined by the HR-diagram for a majority of cases.


\section{Conclusion}

Our evaluation methodology shows that our proposed method allows for codes like {\tt POSYDON} to produce accurate approximations for complete binary evolution tracks providing the time dependence of dozens of physical properties as a function time for interacting binaries, something that has not been possible so far when including the best possible physical treatment accounting for the full stellar structure and evolution. Such a tool will enable astrophysical population studies that require knowledge of such time series, e.g., studies of X-ray binaries through their accretion phase (persistent or transient, calculation of X-ray luminosities and their physical properties distributions as sampled by observatories), of radio pulsar evolution and the spin-up of neutron stars in binaries, etc. Our method can also be further extended to other areas of astrophysics where generation of time series from a pre-computed set of expensive simulations is needed (as well as domains outside astrophysics). 

It is not uncommon for the timescale variance problem to exist in problems that involve simulations where the initial conditions are varied. Our method addresses the timescale variance problem by finding significant changes in signal value to create an alignment between signals which are irregularly sampled and the same methodology could be extended to problems exhibiting the same characteristics. One limitation of our proposed method assumes morphologically homogeneous signals which can be achieved by effective classification. In our work we employ the classification by mass transfer presented in \cite{fragos_posydon_2023}, but further work needs to be done to achieve classification by morphology which would allow for further optimization of important hyperparameters. Furthermore, in the changepoint literature there exist more sophisticated changepoint algorithms \citep{adams_bayesian_2007} which do not require a preset number of changepoints and can be used with our algorithm with minimal modifications. It is our suspicion that effectively employing these algorithms would lead to a slight increase in performance as well as a way to cluster signals by their morphology depending on the number of changepoints needed to represent each signal.

Our method can also be thought of as an interpolation in a latent space. A common idea in signal processing is that of latent space interpolation. Typically a latent space is used to refer to a transformation of some a dataset such that each signal in the latent space is represented by the features or key identifying characteristics of the signal in the original space. The latent space is also usually lower in dimensionality than the original space. For example, the points in time where our simulations exhibit fast changes in signal values correspond can be seen as the underlying features which characterize the signal. 

Therefore, we can see the application of our changepoint treatment as a feature extraction or a mapping into a latent space where interpolation is performed. This makes the changepoint algorithm what is known as an encoder while the linear interpolation which is used to reconstruct our interpolated track is known as the decoder. This formulation of our method points to the use of deep learning generative techniques in which the encoder and decoder are learned through gradient descent methods instead of predefined algorithms. However, given the difficulties in evaluating the performance which we outlined, we found that constructing a loss function that effectively learns can be challenging.

\section{Acknowledgements}

    This work is supported primarily by two sources: the Gordon and Betty Moore Foundation (PI Kalogera, grant awards GBMF8477 and GBMF12341) and the Swiss National Science Foundation (PI Tassos Fragos, project numbers PP00P2\_211006 and CRSII5\_213497).  
    P.M.S, U.D, A.K, V.K, E.T, K.A.R., and M.S. were supported by the project numbers GBMF8477 and GBMF12341.
    J.J.A.~acknowledges support for Program number (JWST-AR-04369.001-A) provided through a grant from the STScI under NASA contract NAS5-03127. 
    S.S.B., M.U.K., and Z.X. were supported by the project number PP00P2\_211006. 
    S.S.B. and M.M.B. were supported by the project number CRSII5\_213497. 
    M.M.B. is also supported by the Boninchi Foundation and the Swiss Government Excellence Scholarship. 
    K.K. is supported by a fellowship program at the Institute of Space Sciences (ICE-CSIC) funded by the program Unidad de Excelencia Mar\'ia de Maeztu CEX2020-001058-M. 
    Z.X. acknowledges support from the Chinese Scholarship Council (CSC). 
    E.Z. acknowledges support from the Hellenic Foundation for Research and Innovation (H.F.R.I.) under the “3rd Call for H.F.R.I. Research Projects to support Post-Doctoral Researchers” (Project No: 7933). 
    The computations were performed at Northwestern University on the Trident computer cluster (funded by the GBMF8477 award) and at the University of Geneva on the Yggdrasil computer cluster. This research was partly supported by the computational resources and staff contributions provided for the Quest high-performance computing facility at Northwestern University, jointly supported by the Office of the Provost, the Office for Research and Northwestern University Information Technology.

\bibliography{main}

\begin{thebibliography}{}
\expandafter\ifx\csname natexlab\endcsname\relax\def\natexlab#1{#1}\fi
\providecommand{\url}[1]{\href{#1}{#1}}
\providecommand{\dodoi}[1]{doi:~\href{http://doi.org/#1}{\nolinkurl{#1}}}
\providecommand{\doeprint}[1]{\href{http://ascl.net/#1}{\nolinkurl{http://ascl.net/#1}}}
\providecommand{\doarXiv}[1]{\href{https://arxiv.org/abs/#1}{\nolinkurl{https://arxiv.org/abs/#1}}}

\bibitem[{Adams \& MacKay(2007)}]{adams_bayesian_2007}
Adams, R.~P., \& MacKay, D. J.~C. 2007, Bayesian {Online} {Changepoint} {Detection},  arXiv.
\newblock \url{http://arxiv.org/abs/0710.3742}

\bibitem[{Arora {et~al.}(2016)Arora, Deepali, \& Varshney}]{arora_analysis_2016}
Arora, P., Deepali, \& Varshney, S. 2016, Procedia Computer Science, 78, 507, \dodoi{10.1016/j.procs.2016.02.095}

\bibitem[{Barber {et~al.}(1996)Barber, Dobkin, \& Huhdanpaa}]{barber_quickhull_1996}
Barber, C.~B., Dobkin, D.~P., \& Huhdanpaa, H. 1996, ACM Transactions on Mathematical Software, 22, 469, \dodoi{10.1145/235815.235821}

\bibitem[{{Belczynski} {et~al.}(2008){Belczynski}, {Kalogera}, {Rasio}, {Taam}, {Zezas}, {Bulik}, {Maccarone}, \& {Ivanova}}]{2008ApJS..174..223B}
{Belczynski}, K., {Kalogera}, V., {Rasio}, F.~A., {et~al.} 2008, \apjs, 174, 223, \dodoi{10.1086/521026}

\bibitem[{Berrut \& Trefethen(2004)}]{berrut_barycentric_2004}
Berrut, J.-P., \& Trefethen, L.~N. 2004, SIAM Review, 46, 501, \dodoi{10.1137/S0036144502417715}

\bibitem[{{Breivik} {et~al.}(2020){Breivik}, {Coughlin}, {Zevin}, {Rodriguez}, {Kremer}, {Ye}, {Andrews}, {Kurkowski}, {Digman}, {Larson}, \& {Rasio}}]{2020ApJ...898...71B}
{Breivik}, K., {Coughlin}, S., {Zevin}, M., {et~al.} 2020, \apj, 898, 71, \dodoi{10.3847/1538-4357/ab9d85}

\bibitem[{Das {et~al.}(2024)Das, Kong, Sen, \& Zhou}]{das_decoder-only_2024}
Das, A., Kong, W., Sen, R., \& Zhou, Y. 2024, A decoder-only foundation model for time-series forecasting.
\newblock \url{http://arxiv.org/abs/2310.10688}

\bibitem[{Dotter(2016)}]{dotter_mesa_2016}
Dotter, A. 2016, The Astrophysical Journal Supplement Series, 222, 8, \dodoi{10.3847/0067-0049/222/1/8}

\bibitem[{{Eldridge} {et~al.}(2017){Eldridge}, {Stanway}, {Xiao}, {McClelland}, {Taylor}, {Ng}, {Greis}, \& {Bray}}]{2017PASA...34...58E}
{Eldridge}, J.~J., {Stanway}, E.~R., {Xiao}, L., {et~al.} 2017, \pasa, 34, e058, \dodoi{10.1017/pasa.2017.51}

\bibitem[{Fragos {et~al.}(2023)Fragos, Andrews, Bavera, Berry, Coughlin, Dotter, Giri, Kalogera, Katsaggelos, Kovlakas, Lalvani, Misra, Srivastava, Qin, Rocha, Roman-Garza, Serra, Stahle, Sun, Teng, Trajcevski, Tran, Xing, Zapartas, \& Zevin}]{fragos_posydon_2023}
Fragos, T., Andrews, J.~J., Bavera, S.~S., {et~al.} 2023, The Astrophysical Journal Supplement Series, 264, 45, \dodoi{10.3847/1538-4365/ac90c1}

\bibitem[{{Giacobbo} \& {Mapelli}(2018)}]{2018MNRAS.480.2011G}
{Giacobbo}, N., \& {Mapelli}, M. 2018, \mnras, 480, 2011, \dodoi{10.1093/mnras/sty1999}

\bibitem[{Hertzsprung(1905)}]{hertzsprung_zur_1905}
Hertzsprung, E. 1905, Zeitschrift für wissenschaftliche Photographie, Photochemie und Photophysik, 3, 429

\bibitem[{Hochreiter \& Schmidhuber(1997)}]{hochreiter_long_1997}
Hochreiter, S., \& Schmidhuber, J. 1997, Neural Computation, 9, 1735

\bibitem[{{Hurley} {et~al.}(2000){Hurley}, {Pols}, \& {Tout}}]{2000MNRAS.315..543H}
{Hurley}, J.~R., {Pols}, O.~R., \& {Tout}, C.~A. 2000, \mnras, 315, 543, \dodoi{10.1046/j.1365-8711.2000.03426.x}

\bibitem[{{Hurley} {et~al.}(2002){Hurley}, {Tout}, \& {Pols}}]{2002MNRAS.329..897H}
{Hurley}, J.~R., {Tout}, C.~A., \& {Pols}, O.~R. 2002, \mnras, 329, 897, \dodoi{10.1046/j.1365-8711.2002.05038.x}

\bibitem[{Jenkins(2014)}]{kenett_autoregressiveintegrated_2014}
Jenkins, G.~M. 2014, in Wiley {StatsRef}: {Statistics} {Reference} {Online}, 1st edn., ed. R.~S. Kenett, N.~T. Longford, W.~W. Piegorsch, \& F.~Ruggeri (Wiley), \dodoi{10.1002/9781118445112.stat03472}

\bibitem[{{Jermyn} {et~al.}(2023){Jermyn}, {Bauer}, {Schwab}, {Farmer}, {Ball}, {Bellinger}, {Dotter}, {Joyce}, {Marchant}, {Mombarg}, {Wolf}, {Sunny Wong}, {Cinquegrana}, {Farrell}, {Smolec}, {Thoul}, {Cantiello}, {Herwig}, {Toloza}, {Bildsten}, {Townsend}, \& {Timmes}}]{2023ApJS..265...15J}
{Jermyn}, A.~S., {Bauer}, E.~B., {Schwab}, J., {et~al.} 2023, \apjs, 265, 15, \dodoi{10.3847/1538-4365/acae8d}

\bibitem[{Kang {et~al.}(2020)Kang, Hyndman, \& Li}]{kang_gratis_2020}
Kang, Y., Hyndman, R.~J., \& Li, F. 2020, Statistical Analysis and Data Mining: The ASA Data Science Journal, 13, 354, \dodoi{10.1002/sam.11461}

\bibitem[{{Kruckow} {et~al.}(2018){Kruckow}, {Tauris}, {Langer}, {Kramer}, \& {Izzard}}]{2018MNRAS.481.1908K}
{Kruckow}, M.~U., {Tauris}, T.~M., {Langer}, N., {Kramer}, M., \& {Izzard}, R.~G. 2018, \mnras, 481, 1908, \dodoi{10.1093/mnras/sty2190}

\bibitem[{Li {et~al.}(2023)Li, Yu, \& Principe}]{li_causal_2023}
Li, H., Yu, S., \& Principe, J. 2023, Causal {Recurrent} {Variational} {Autoencoder} for {Medical} {Time} {Series} {Generation}.
\newblock \url{http://arxiv.org/abs/2301.06574}

\bibitem[{Liao {et~al.}(2023)Liao, Ni, Szpruch, Wiese, Sabate-Vidales, \& Xiao}]{liao_conditional_2023}
Liao, S., Ni, H., Szpruch, L., {et~al.} 2023, Conditional {Sig}-{Wasserstein} {GANs} for {Time} {Series} {Generation},  arXiv.
\newblock \url{http://arxiv.org/abs/2006.05421}

\bibitem[{Madane {et~al.}(2022)Madane, Dilmi, Forest, Azzag, Lebbah, \& Lacaille}]{madane_transformer-based_2022}
Madane, A., Dilmi, M.-d., Forest, F., {et~al.} 2022, Transformer-based conditional generative adversarial network for multivariate time series generation.
\newblock \url{http://arxiv.org/abs/2210.02089}

\bibitem[{Maltsev {et~al.}(2024)Maltsev, Schneider, Roepke, Jordan, Qadir, Kerzendorf, Riedmiller, \& van~der Smagt}]{maltsev_scalable_2024}
Maltsev, K., Schneider, F. R.~N., Roepke, F.~K., {et~al.} 2024, Astronomy \& Astrophysics, 681, A86, \dodoi{10.1051/0004-6361/202347118}

\bibitem[{{Nelemans} {et~al.}(2001){Nelemans}, {Yungelson}, {Portegies Zwart}, \& {Verbunt}}]{2001A&A...365..491N}
{Nelemans}, G., {Yungelson}, L.~R., {Portegies Zwart}, S.~F., \& {Verbunt}, F. 2001, \aap, 365, 491, \dodoi{10.1051/0004-6361:20000147}

\bibitem[{Nygaard {et~al.}(2001)Nygaard, Melnikov, \& Katsaggelos}]{900246}
Nygaard, R., Melnikov, G., \& Katsaggelos, A. 2001, IEEE Transactions on Biomedical Engineering, 48, 28, \dodoi{10.1109/10.900246}

\bibitem[{Oppenheim(1999)}]{oppenheim1999discrete}
Oppenheim, A.~V. 1999, Discrete-time signal processing (Pearson Education India)

\bibitem[{{Paxton} {et~al.}(2011){Paxton}, {Bildsten}, {Dotter}, {Herwig}, {Lesaffre}, \& {Timmes}}]{2011ApJS..192....3P}
{Paxton}, B., {Bildsten}, L., {Dotter}, A., {et~al.} 2011, \apjs, 192, 3, \dodoi{10.1088/0067-0049/192/1/3}

\bibitem[{{Paxton} {et~al.}(2013){Paxton}, {Cantiello}, {Arras}, {Bildsten}, {Brown}, {Dotter}, {Mankovich}, {Montgomery}, {Stello}, {Timmes}, \& {Townsend}}]{2013ApJS..208....4P}
{Paxton}, B., {Cantiello}, M., {Arras}, P., {et~al.} 2013, \apjs, 208, 4, \dodoi{10.1088/0067-0049/208/1/4}

\bibitem[{{Paxton} {et~al.}(2015){Paxton}, {Marchant}, {Schwab}, {Bauer}, {Bildsten}, {Cantiello}, {Dessart}, {Farmer}, {Hu}, {Langer}, {Townsend}, {Townsley}, \& {Timmes}}]{2015ApJS..220...15P}
{Paxton}, B., {Marchant}, P., {Schwab}, J., {et~al.} 2015, \apjs, 220, 15, \dodoi{10.1088/0067-0049/220/1/15}

\bibitem[{{Paxton} {et~al.}(2018){Paxton}, {Schwab}, {Bauer}, {Bildsten}, {Blinnikov}, {Duffell}, {Farmer}, {Goldberg}, {Marchant}, {Sorokina}, {Thoul}, {Townsend}, \& {Timmes}}]{2018ApJS..234...34P}
{Paxton}, B., {Schwab}, J., {Bauer}, E.~B., {et~al.} 2018, \apjs, 234, 34, \dodoi{10.3847/1538-4365/aaa5a8}

\bibitem[{{Paxton} {et~al.}(2019){Paxton}, {Smolec}, {Schwab}, {Gautschy}, {Bildsten}, {Cantiello}, {Dotter}, {Farmer}, {Goldberg}, {Jermyn}, {Kanbur}, {Marchant}, {Thoul}, {Townsend}, {Wolf}, {Zhang}, \& {Timmes}}]{2019ApJS..243...10P}
{Paxton}, B., {Smolec}, R., {Schwab}, J., {et~al.} 2019, \apjs, 243, 10, \dodoi{10.3847/1538-4365/ab2241}

\bibitem[{Paxton {et~al.}(2019)Paxton, Smolec, Schwab, Gautschy, Bildsten, Cantiello, Dotter, Farmer, Goldberg, Jermyn, Kanbur, Marchant, Thoul, Townsend, Wolf, Zhang, \& Timmes}]{paxton_modules_2019}
Paxton, B., Smolec, R., Schwab, J., {et~al.} 2019, The Astrophysical Journal Supplement Series, 243, 10, \dodoi{10.3847/1538-4365/ab2241}

\bibitem[{{Pols} {et~al.}(1995){Pols}, {Tout}, {Eggleton}, \& {Han}}]{1995MNRAS.274..964P}
{Pols}, O.~R., {Tout}, C.~A., {Eggleton}, P.~P., \& {Han}, Z. 1995, \mnras, 274, 964, \dodoi{10.1093/mnras/274.3.964}

\bibitem[{{Riley} {et~al.}(2022){Riley}, {Agrawal}, {Barrett}, {Boyett}, {Broekgaarden}, {Chattopadhyay}, {Gaebel}, {Gittins}, {Hirai}, {Howitt}, \& et~al.}]{2021arXiv210910352T}
{Riley}, J., {Agrawal}, P., {Barrett}, J.~W., {et~al.} 2022, \apjs, 258, 34, \dodoi{10.3847/1538-4365/ac416c}

\bibitem[{Rocha {et~al.}(2022)Rocha, Andrews, Berry, Doctor, Katsaggelos, Pérez, Marchant, Kalogera, Coughlin, Bavera, Dotter, Fragos, Kovlakas, Misra, Xing, \& Zapartas}]{rocha_active_2022}
Rocha, K.~A., Andrews, J.~J., Berry, C. P.~L., {et~al.} 2022, The Astrophysical Journal, 938, 64, \dodoi{10.3847/1538-4357/ac8b05}

\bibitem[{Russel(1919)}]{russel_problems_1919}
Russel, H.~N. 1919, Proceedings of the National Academy of Sciences, 5, 391

\bibitem[{{Spera} {et~al.}(2019){Spera}, {Mapelli}, {Giacobbo}, {Trani}, {Bressan}, \& {Costa}}]{2019MNRAS.485..889S}
{Spera}, M., {Mapelli}, M., {Giacobbo}, N., {et~al.} 2019, \mnras, 485, 889, \dodoi{10.1093/mnras/stz359}

\bibitem[{{Stanway} \& {Eldridge}(2018)}]{2018MNRAS.479...75S}
{Stanway}, E.~R., \& {Eldridge}, J.~J. 2018, \mnras, 479, 75, \dodoi{10.1093/mnras/sty1353}

\bibitem[{{Toonen} {et~al.}(2012){Toonen}, {Nelemans}, \& {Portegies Zwart}}]{2012A&A...546A..70T}
{Toonen}, S., {Nelemans}, G., \& {Portegies Zwart}, S. 2012, \aap, 546, A70, \dodoi{10.1051/0004-6361/201218966}

\bibitem[{{Tout} {et~al.}(1997){Tout}, {Aarseth}, {Pols}, \& {Eggleton}}]{1997MNRAS.291..732T}
{Tout}, C.~A., {Aarseth}, S.~J., {Pols}, O.~R., \& {Eggleton}, P.~P. 1997, \mnras, 291, 732, \dodoi{10.1093/mnras/291.4.732}

\bibitem[{Vaswani {et~al.}(2023)Vaswani, Shazeer, Parmar, Uszkoreit, Jones, Gomez, Kaiser, \& Polosukhin}]{vaswani_attention_2023}
Vaswani, A., Shazeer, N., Parmar, N., {et~al.} 2023, Attention {Is} {All} {You} {Need},  arXiv.
\newblock \url{http://arxiv.org/abs/1706.03762}

\bibitem[{Yuan \& Qiao(2024)}]{yuan_diffusion-ts_2024}
Yuan, X., \& Qiao, Y. 2024, Diffusion-{TS}: {Interpretable} {Diffusion} for {General} {Time} {Series} {Generation},  arXiv.
\newblock \url{http://arxiv.org/abs/2403.01742}

\end{thebibliography}
\bibliographystyle{aasjournal}



\end{document}